\documentclass[12pt,twocolumn]{article}

\usepackage[margin=1in,top=1.2in,bottom=1.5in]{geometry}
\usepackage{amsmath}
\usepackage{amssymb}
\usepackage{booktabs}
\usepackage{siunitx}
\usepackage{multirow}
\usepackage{graphicx}
\usepackage{float}
\usepackage{placeins}
\usepackage{url}
\usepackage{authblk}
\usepackage{natbib}
\usepackage[hidelinks]{hyperref}

\newcommand{\keywords}[1]{%
  \begin{center}
    \begin{minipage}{0.82\textwidth}
      \small\noindent\textbf{Keywords:} \textit{#1}
    \end{minipage}
  \end{center}
  \vspace{1em}%
}

\renewenvironment{abstract}{%
  \vspace{1.5em}%
  \begin{center}
    \begin{minipage}{0.82\textwidth}
      \noindent\rule{\linewidth}{0.6pt}
      \vspace{0.4em}
      \begin{center}{\normalsize\bfseries Abstract}\end{center}
      \vspace{0.2em}
      \small
}{%
      \par\vspace{0.5em}
      \noindent\rule{\linewidth}{0.6pt}
    \end{minipage}
  \end{center}
  \vspace{0.5em}%
}

\title{Probing heartbeat oscillations from the black hole X-ray binary GRS 1915+105 using spectral-timing analysis}

\author[1]{Karan Akbari\thanks{Corresponding author: \href{mailto:karanakbari14@gmail.com}{\texttt{karanakbari14@gmail.com}}; ORCID: 0009-0005-0550-4018}}
\author[1]{Chintan Patel\thanks{\href{mailto:chintapatelhea@gmail.com}{\texttt{chintapatelhea@gmail.com}}; ORCID: 0009-0006-8622-5471}}
\author[2]{Sayantan Bhattacharya\thanks{\href{mailto:sayantan.bhattacharya@tifr.res.in}{\texttt{sayantan.bhattacharya@tifr.res.in}}; ORCID: 0000-0001-8572-8241}}
\author[2]{Sudip Bhattacharyya\thanks{\href{mailto:sudip@tifr.res.in}{\texttt{sudip@tifr.res.in}}; ORCID: 0000-0002-6351-5808}}
\author[1]{Manojendu Choudhury\thanks{\href{mailto:manojendu.choudhury@xaviers.edu}{\texttt{manojendu.choudhury@xaviers.edu}}; ORCID: 0000-0002-3021-6190}}

\affil[1]{St.\ Xavier's College, Mumbai, India}
\affil[2]{Department of Astronomy and Astrophysics, Tata Institute of Fundamental Research, 1 Homi Bhabha Road, Colaba, Mumbai, India}

\newcommand{\new}[1]{#1}

\begin{document}
\onecolumn
\setcounter{topnumber}{4}
\setcounter{bottomnumber}{2}
\setcounter{totalnumber}{6}
\renewcommand{\topfraction}{0.95}
\renewcommand{\bottomfraction}{0.5}
\renewcommand{\textfraction}{0.05}
\renewcommand{\floatpagefraction}{0.99}
\renewcommand{\dbltopfraction}{0.95}
\renewcommand{\dblfloatpagefraction}{0.99}

\maketitle

\begin{abstract}

GRS~1915+105 is a black hole X-ray binary whose $\rho$-class (``heartbeat'') oscillations ($\sim$50--100\,s) are attributed to radiation-pressure instabilities in the inner accretion disk at near-Eddington luminosities. We present a phase-resolved spectral and timing analysis of 24 \textit{Swift} XRT observations (1--10\,keV) and broadband \textit{AstroSat} SXT+LAXPC data (0.8--30\,keV), dividing each cycle into five phases. The narrow-band XRT fits show an apparent anti-correlation between the inner disk temperature ($T_{\rm in}\sim1.7$--$1.5$\,keV) and apparent radius ($R_{\rm in}\sim22$--$38$\,km) across the cycle. \new{The broadband \textit{AstroSat} fits, however, are statistically consistent with a constant disk temperature: a joint fit with $T_{\rm in}$ tied across all five phases gives $T_{\rm in}=1.275\pm0.020$\,keV ($\chi^2_\nu=1.003$; $\Delta\chi^2=+4.3$ for 4 added constraints), whereas tying the disk normalization as well is rejected ($\Delta\chi^2=+175.9$), leaving a $\sim$20\% variation in apparent $R_{\rm in}$ ($18.1\pm0.7$ to $21.9\pm0.7$\,km).} The coronal electron temperature rises from $\sim$6 to $\sim$14.5\,keV approaching the burst, with the photon index tracking it. \new{We attribute the larger XRT disk swings to its limited bandpass, where coronal Comptonization is unconstrained and the disk parameters absorb coronal variability; the dominant variability is therefore coronal, consistent with \citet{Vadawale2001}, and the residual $R_{\rm in}$ change is plausibly a color-correction effect \citep{Zoghbi2016}.} Hardness--intensity and color--color diagrams show clear spectral hysteresis. Our broadband coverage provides a phase-resolved test of disk constancy and favors coronal evolution as the driver of the spectral variability across the $\rho$ cycle.

\end{abstract}


\twocolumn
\section{Introduction}
 
\begin{figure*}[tbp]
    \centering
    \includegraphics[width=\textwidth]{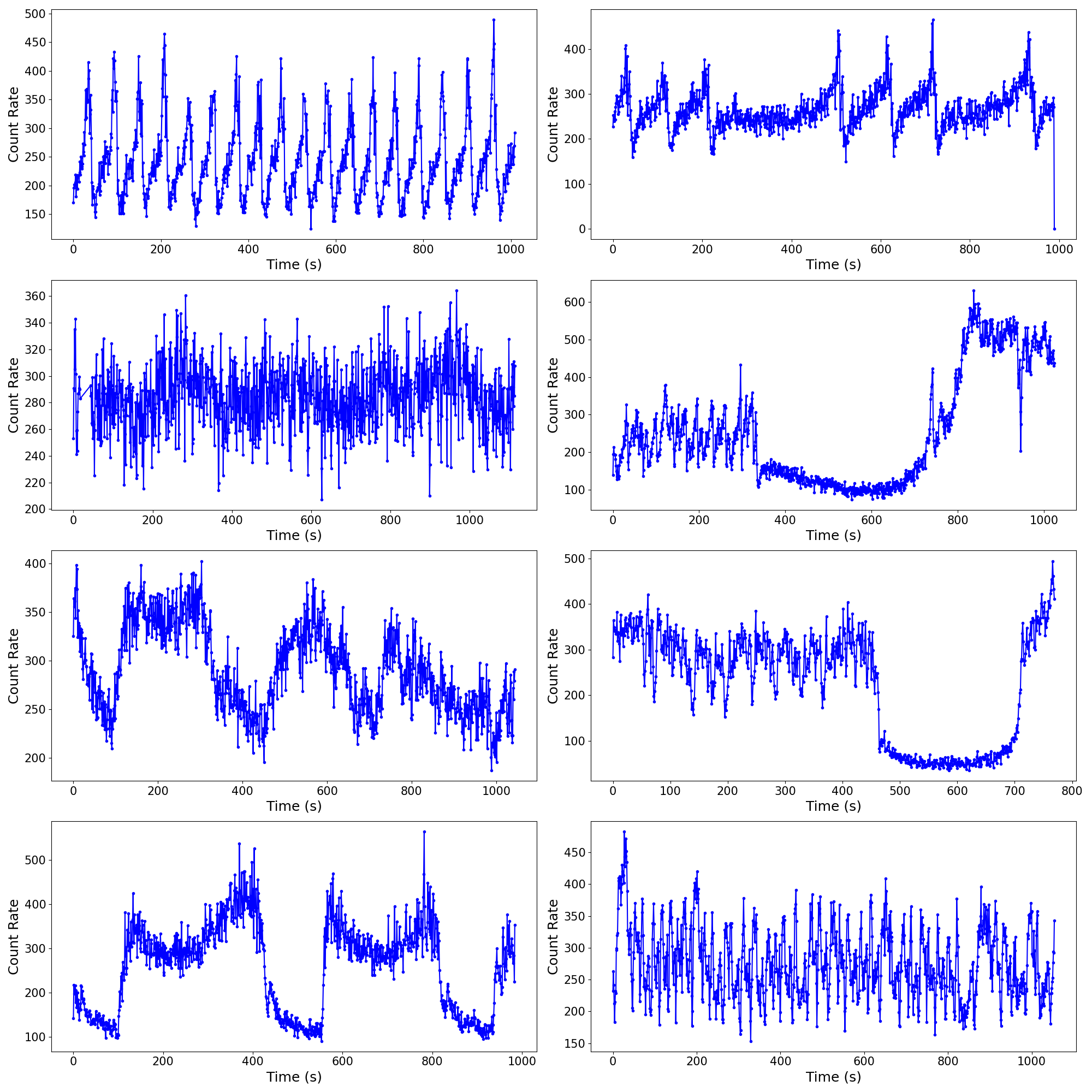}
    \caption{Representative one-second-binned \textit{Swift} XRT light curves of eight variability classes identified in our survey. From top left to bottom right: classes $\rho$, $\nu$, $\chi$, $\beta$, $\delta$, $\lambda$, $\theta$ and $\mu$ (as defined by \citet{belloni2000variability}). Each panel spans 1000 s and is plotted on the same flux scale to highlight the distinct morphology of steady emission, quasi-periodic bursts, dip-flare cycles and erratic flickering exhibited by GRS~1915+105. This paper focuses on the $\rho$-class; see Section~\ref{sec:data}.}
    \label{fig:variability_classes}
\end{figure*}

GRS~1915+105 is a well-studied black hole X-ray binary in the Galaxy, distinguished by its persistent activity, superluminal jets, and complex X-ray variability \citep{castro1992discovery, mirabel1994superluminal}. Located at $\sim$8.6\,kpc and harboring a $12.4 \pm 2.0\,M_\odot$ black hole \citep{reid2014grs}, it has near-maximal spin $a_* > 0.98$ \citep{McClintock2006, Miller2013}. At least twelve distinct variability classes have been identified \citep{belloni2000variability, remillard2006xrb, done2007xrb}; Figure~\ref{fig:variability_classes} illustrates eight classes from our \textit{Swift} XRT survey.
 
Among these, the $\rho$-class (``heartbeat'') oscillation is characterized by quasi-periodic flux variations with typical periods of $\sim$50--100\,s, believed to arise from radiation-pressure-driven instabilities in the inner accretion disk \citep{Neilsen_2011, Zoghbi2016}, making it a valuable probe of near-Eddington accretion physics.
 
At high accretion rates approaching the Eddington limit, radiation pressure can exceed gas pressure support, producing a thermally and viscously unstable disk \citep{LightmanEardley1974, ShakuraSunyaev1976, abramowicz1988slim}. This instability manifests as a cyclic sequence: radiation pressure inflates the inner disk outward, local temperature drops, support eventually fails, infalling material heats as it reaches the innermost regions, and a burst is produced before the disk resets. This mechanism has been proposed to explain the $\rho$-class oscillations in GRS~1915+105 \citep{Neilsen_2011}.
 
\citet{Neilsen_2011} provided the first detailed phase-resolved spectral analysis of the $\rho$ variability using \textit{Chandra} HETGS and \textit{RXTE}, establishing a causal link between the heartbeat oscillations and disk-wind interaction. \citet{Zoghbi2016} subsequently used simultaneous \textit{NuSTAR}/\textit{Chandra} observations to show that while the continuum inferred disk inner radius varies through the cycle, the reflection-inferred radius remains nearly constant, implicating changes in the spectral hardening factor. We note that while ``heartbeat'' and $\rho$-class are often used interchangeably, \citet{Rawat_2019} identified timing differences between the two, with heartbeat periods of 100--150\,s compared to the canonical $\rho$-class periods of $\sim$50--100\,s. Despite these advances, broadband multi-instrument phase-resolved studies constraining both the disk and coronal continuum have remained limited.
 
In this paper, we present a phase-resolved spectral and timing analysis of the $\rho$-class variability in GRS~1915+105 using \textit{Swift} XRT and \textit{AstroSat} LAXPC/SXT. Our primary goals are to characterize the systematic evolution of accretion disk parameters, inner disk temperature $T_{\rm in}$ and apparent inner radius $R_{\rm in}$, across the five phases of the $\rho$ cycle through phase-resolved spectroscopy, to investigate spectral state transitions via Hardness--Intensity and Color--Color Diagrams, and to provide new multi-instrument, broadband observational constraints on the radiation-pressure instability model. Our analysis uses the complementary capabilities of these instruments: \textit{Swift} XRT provides extensive temporal coverage across 24 observations from 2014 to 2016, while \textit{AstroSat}'s broadband capability (0.3--80\,keV) simultaneously constrains both the thermal disk and Comptonized coronal continuum, which is not achievable from soft X-ray data alone.
 
This paper is organized as follows: Section~\ref{sec:data} describes the observations and data reduction procedures for both \textit{Swift} XRT and \textit{AstroSat} LAXPC/SXT. Section~\ref{sec:phase} details our phase-resolved analysis methodology. Section~\ref{sec:results} presents the spectral fitting results and phase-resolved parameter evolution. Section~\ref{sec:discussion} discusses the physical interpretation of our findings in the context of radiation-pressure instabilities. Finally, Section~\ref{sec:conclusions} summarizes our main conclusions.
 
\section{Observations and Data Reduction}
\label{sec:data}
 
\subsection{\textit{Swift} XRT Observations and Data Reduction}
All available \textit{Swift} XRT observations of GRS\,1915+105 were retrieved from the HEASARC archive\footnote{\url{https://heasarc.gsfc.nasa.gov}}. In total, 702 observations spanning 2005--2024 (MJD 53300--59900) were downloaded. Because our analysis focuses on timing and spectroscopy rather than imaging, and Photon Counting (PC) mode data were sparse and of lower quality for this bright source, all processing was restricted to Windowed Timing (WT) mode. Event files were processed with \texttt{xrtpipeline} in HEASoft v6.29\footnote{\url{https://heasarc.gsfc.nasa.gov/docs/software/lheasoft/}} and barycenter-corrected using \texttt{barycorr}. Source spectra and light curves were extracted using 40\,arcsec circular apertures with annular background regions.
 
Light curves were binned at 1\,s and inspected to identify the $\sim$100\,s flares characteristic of the $\rho$-class, confirmed via power-spectral analysis \citep{Neilsen_2011, belloni2000variability}. Twenty-four observations from 2014--2016 (MJD 56600--57500) show the $\rho$-class (``heartbeat'') oscillation. Spectra were combined on a phase-by-phase basis and grouped to a minimum of 20 counts per bin prior to spectral fitting. All XRT spectra were analyzed over the 1.0--10.0\,keV band, where the WT mode response is well calibrated.
 
\subsection{\textit{AstroSat} LAXPC and SXT Observations and Data Reduction}
\label{sec:astrosat}
 
All \textit{AstroSat} observations of GRS\,1915+105 were retrieved from the AstroBrowse archive\footnote{\url{https://astrobrowse.issdc.gov.in/astro\_archive/archive/Home.jsp}}. A total of 40 LAXPC and 49 SXT datasets were obtained. $\rho$-class intervals were first identified in the 3--30\,keV LAXPC light curves; only the corresponding SXT observations were retained for joint analysis. LAXPC spectra were used over the 3--30\,keV range and grouped to a minimum of 30 counts per bin.
 
LAXPC data were processed using the LAXPCsoft pipeline\footnote{\url{http://www.tifr.res.in/~astrosat\_laxpc/LaxpcSoft.html}} to produce calibrated event files, good time intervals, and background-subtracted spectra and light curves. Dead-time corrections were applied at the pipeline level. SXT event files from $\rho$-class intervals were merged using the \texttt{SXTMerger} tool\footnote{\url{http://astrosat-ssc.iucaa.in/uploads/threadsPageNew\_SXT.html}}, and spectra were extracted in the 0.3--7\,keV band using standard grade filtering and the appropriate response and ancillary files. The SXT spectra were used over the 0.8--7\,keV range for fitting, with the lower limit set to avoid calibration uncertainties below 0.8\,keV. Light curves were binned at 1\,s for both instruments.
 
Two LAXPC observations taken in 2017 (MJD 57856, 57857) show the $\rho$-class oscillation. SXT coverage during these observations was shorter than the full LAXPC exposure, meaning the phase-combined SXT and LAXPC spectra are drawn from partially different subsets of heartbeat cycles. The cross-normalization constant in our joint fits accounts for mean flux offsets between instruments but cannot correct for phase-coherence differences arising from this non-simultaneous coverage.

\begin{table*}[tbp]
    \renewcommand{\arraystretch}{0.9}
    \caption{Selected observations used for $\rho$-class analysis. See Section~\ref{sec:data} for data reduction details.}
    \begin{center}
    \hspace{0cm}
    \begin{tabular}{llccc}
        \toprule
        \textbf{Mission} & \textbf{OBSID} & \textbf{Instrument} & \textbf{Exposure (s)} & \textbf{Start Time (MJD)} \\
        \midrule
        \multirow{24}{*}{Swift} 
        & 00030333020 & XRT & 975 & 56778.21041 \\
        & 00030333021 & XRT & 955 & 56780.73192 \\
        & 00030333022 & XRT & 979 & 56783.00971 \\
        & 00030333023 & XRT & 960 & 56785.14791 \\
        & 00030333026 & XRT & 954 & 56793.81943 \\
        & 00030333028 & XRT & 689 & 56798.21943 \\
        & 00030333029 & XRT & 1079 & 56800.39791 \\
        & 00030333030 & XRT & 940 & 56803.66179 \\
        & 00030333031 & XRT & 1080 & 56805.59513 \\
        & 00030333109 & XRT & 1060 & 57070.06388 \\
        & 00030333110 & XRT & 1099 & 57072.84513 \\
        & 00030333112 & XRT & 1095 & 57080.24513 \\
        & 00030333114 & XRT & 1085 & 57085.76735 \\
        & 00030333118 & XRT & 265 & 57095.40554 \\
        & 00030333119 & XRT & 1050 & 57097.20277 \\
        & 00030333120 & XRT & 999 & 57099.59789 \\
        & 00030333122 & XRT & 895 & 57102.59235 \\
        & 00030333123 & XRT & 945 & 57105.00414 \\
        & 00030333124 & XRT & 895 & 57107.51943 \\
        & 00030333125 & XRT & 955 & 57110.57846 \\
        & 00030333126 & XRT & 1120 & 57112.77498 \\
        & 00030333127 & XRT & 1015 & 57115.04582 \\
        & 00030333138 & XRT & 965 & 57142.46110 \\
        & 00081433001 & XRT & 2008 & 57077.10692 \\
        \midrule
        \multirow{4}{*}{\textit{AstroSat}}
        & \multirow{2}{*}{G07\_028T01\_9000001166} & LAXPC & 9378 & \multirow{2}{*}{57857.0 (2017/04/15)} \\
        &                                           & SXT   & 3604 & \\
        & \multirow{2}{*}{G07\_046T01\_9000001162} & LAXPC & 15269 & \multirow{2}{*}{57856.0 (2017/04/14)} \\
        &                                           & SXT   & 7898 & \\
        \bottomrule
    \end{tabular}
    \end{center}
    \label{tab:observations}
\end{table*}

\section{Phase-Resolved Analysis}
\label{sec:phase}

Phase-resolved analysis of the $\rho$-class oscillations was carried out by segmenting each variability cycle into five discrete phases using barycenter-corrected, one-second binned light curves. The dominant oscillation period was determined via a Lomb--Scargle periodogram in the 20--200\,s range. Local maxima and minima were identified to define phase boundaries: each trough-to-peak interval was divided into three equal phases (Phases~1--3, rise) and each peak-to-trough interval into two equal phases (Phases~4--5, decay). Phase-specific GTIs were then used to extract spectra and light curves for each instrument.

\subsection{Identification of \texorpdfstring{$\rho$}{rho}-Class Intervals}

\begin{figure*}[tbp]
    \centering
    \includegraphics[width=\textwidth]{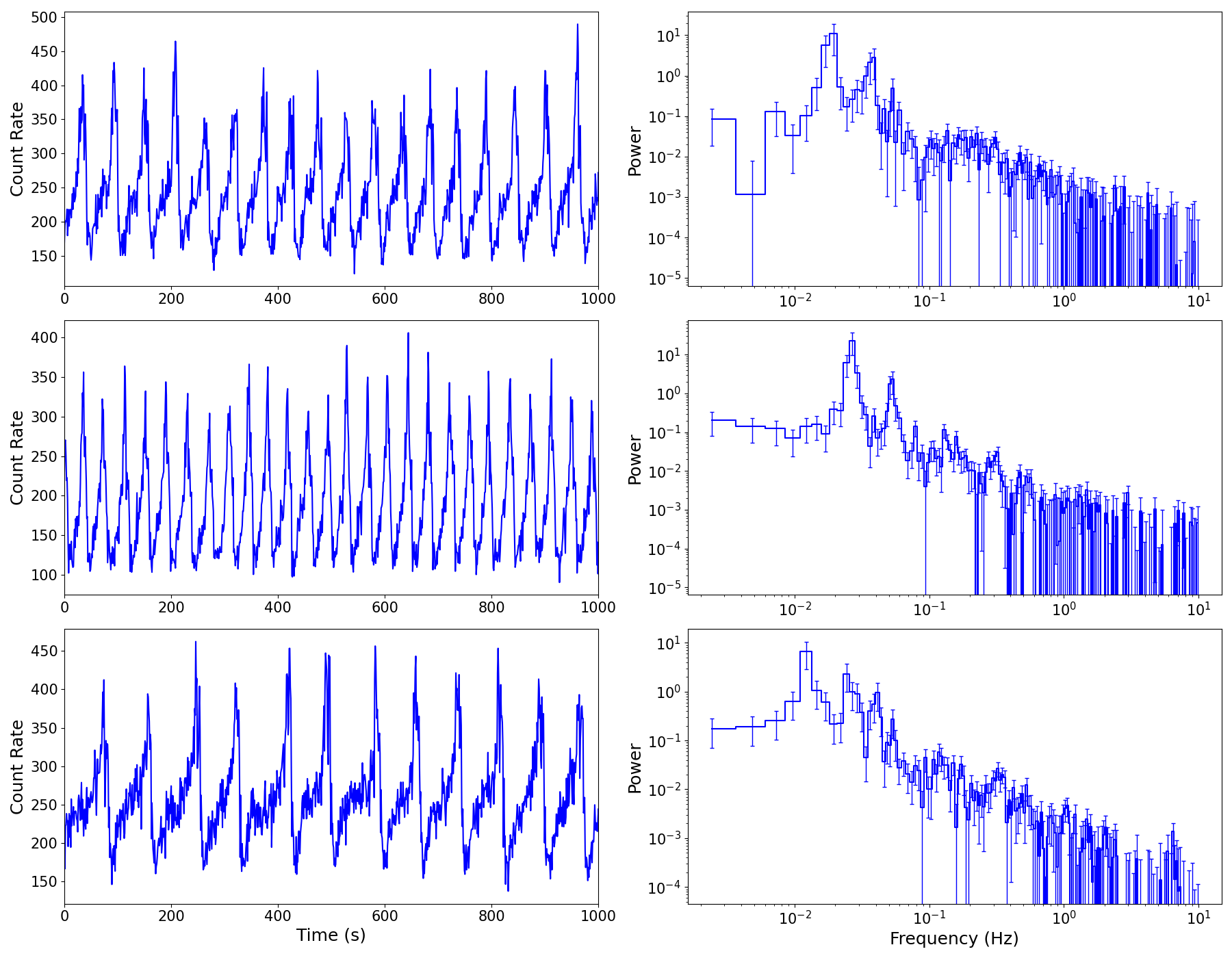}
    \caption{Three examples of $\rho$-class (``heartbeat'') intervals from \textit{Swift} XRT (left column), each paired with its corresponding Leahy-normalized power density spectrum (right column). Light curves are plotted at 1 s resolution over 1000 s segments, illustrating the recurrence of the $\sim$100 s flares. The power spectra show clear peaks at the characteristic oscillation frequency, confirming the quasi-periodic nature of the variability. See Sections~\ref{sec:phase} and \ref{sec:results}.}
    \label{fig:rho_examples}
\end{figure*}

Figure~\ref{fig:rho_examples} presents three representative examples of $\rho$-class variability observed with \textit{Swift} XRT, each paired with its corresponding power density spectrum. The characteristic quasi-periodic oscillations with periods of $\sim$100\,s are visible in the light curves, and the power spectra show prominent peaks confirming the coherent nature of these oscillations. The recurrence and regularity of these ``heartbeat'' flares ensure stability for our phase-resolved analysis.

\subsection{\textit{Swift} XRT Phase-Resolved Analysis}
One-second binned light curves were constructed from the event lists, with errors propagated in quadrature. A Lomb--Scargle periodogram in the 20--200\,s period range identified the dominant oscillation frequency. Peaks and troughs were located with a minimum separation constraint equal to the measured period to suppress spurious detections. Phase boundaries were set at the identified extrema, yielding three rise phases and two decay phases per cycle, and phase-specific GTIs were used to extract spectra and light curves.

\subsection{\textit{AstroSat} LAXPC Phase-Resolved Analysis}
The same phase-segmentation approach was applied to the LAXPC data. One-second binned light curves were produced from the Level-2 event data, and the oscillation period was determined via a Lomb--Scargle periodogram. A sliding-window algorithm identified successive peaks and troughs, partitioning each cycle into three rise phases and two decay phases. Phase-specific GTIs were applied to extract phase-resolved spectra and light curves for each observation.

\begin{figure*}[tbp]
    \centering
    \includegraphics[width=\textwidth]{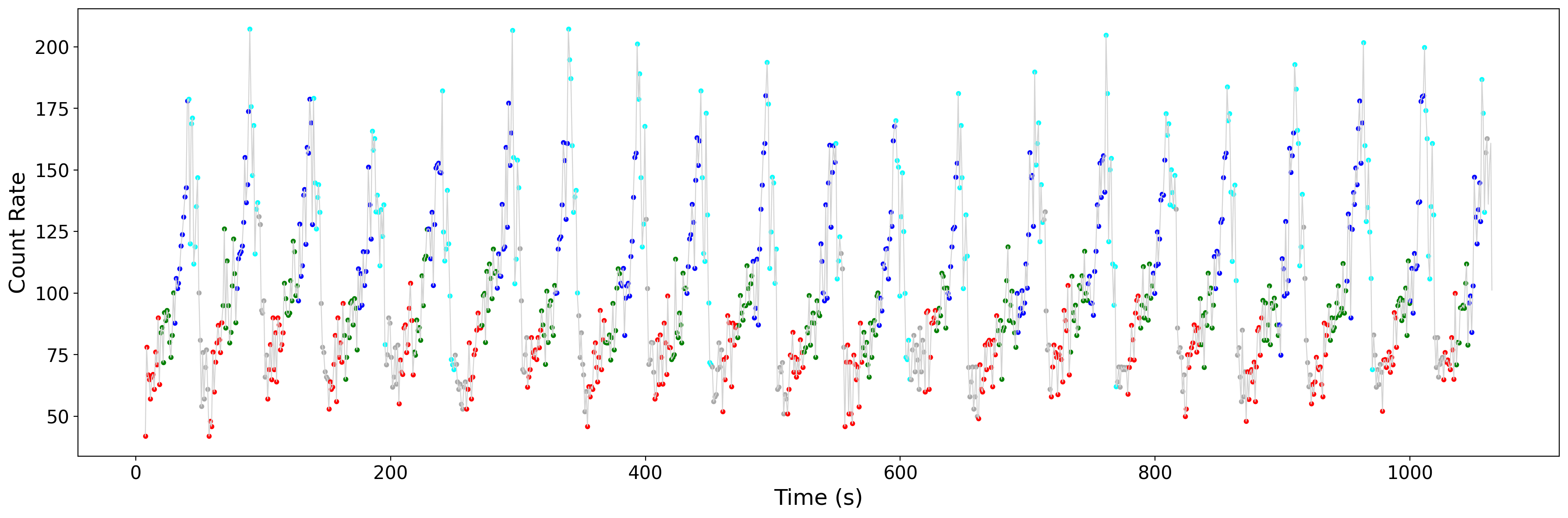}
    \caption{Phase classification for a single $\rho$-class observation (\textit{Swift} XRT OBSID 00030333020). The full 200 s light curve is shown in gray; colored points indicate the five phase bins: phases 1--3 (rise; red, green, blue) and phases 4--5 (decay; cyan, gray). Vertical dashed lines mark the boundaries between phases. This color scheme is used throughout all HID/CCD figures. See Section~\ref{sec:phase}.}
    \label{fig:phase_classification}
\end{figure*}

\begin{figure*}[tbp]
    \centering
    \includegraphics[width=0.48\textwidth]{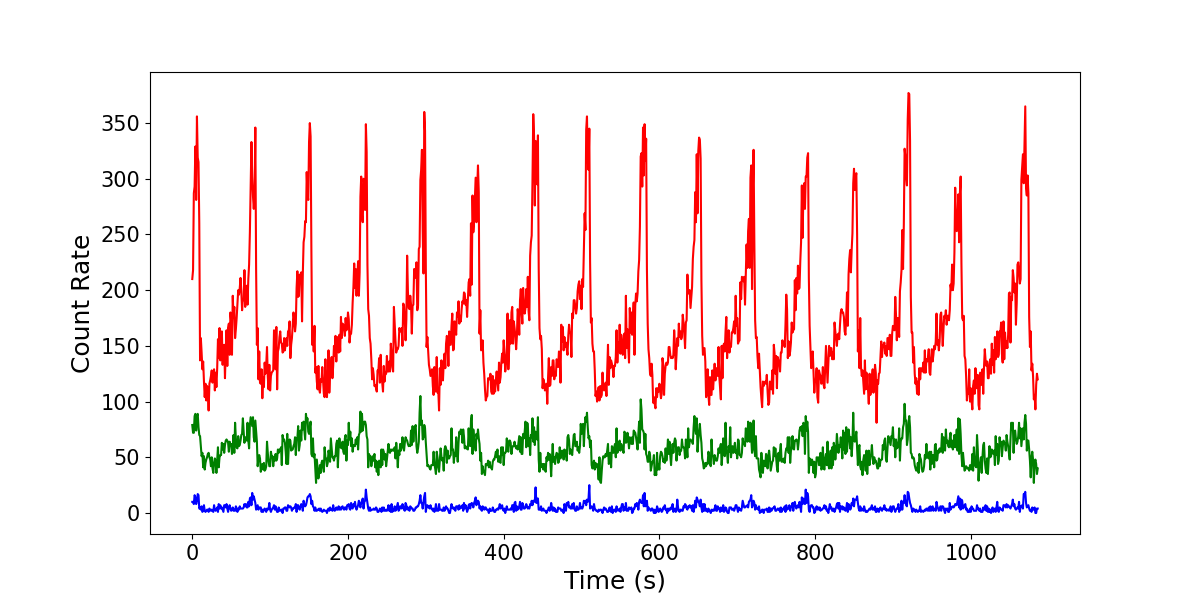}
    \hfill
    \includegraphics[width=0.48\textwidth]{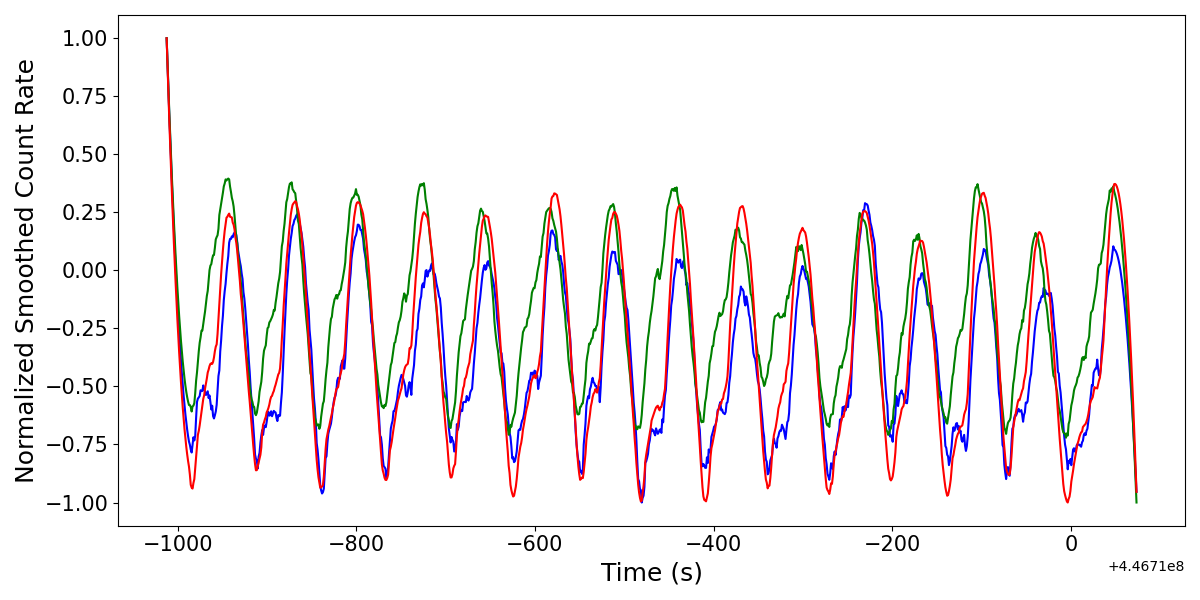}
    \caption{\textit{Left:} Energy-resolved light curves of multiple $\rho$-class cycles from \textit{Swift} XRT (OBSID 00030333112) in three bands: low (0.3--1\,keV, blue), mid (1--3\,keV, green), and high (3--10\,keV, red). The heartbeat pattern recurs with a period of $\sim$100\,s across all bands, with flare amplitude increasing toward higher energies. \textit{Right:} Smoothed and normalized ($-1$ to $+1$) light curves after applying a Savitzky--Golay filter, highlighting the energy-dependent morphology. The high-energy band leads the softer bands by several seconds, consistent with inward-to-outward propagation of the disk variability. See Sections~\ref{sec:phase} and \ref{sec:results} for discussion.}
    \label{fig:energy_resolved_lc}
    \label{fig:smoothed_lc}
\end{figure*}

Figure~\ref{fig:phase_classification} illustrates our phase-segmentation scheme applied to a single $\rho$-class cycle. The oscillation is divided into five phases: three rising phases (Phases~1--3) and two decay phases (Phases~4--5), with phase boundaries determined by the local maxima and minima in the light curve. Figure~\ref{fig:energy_resolved_lc} shows the energy-resolved light curves of the heartbeat oscillations alongside their smoothed, normalized counterparts, highlighting the energy-dependent morphology and the phase offsets between bands.

\noindent Phase-resolved spectra from the 24 \textit{Swift} XRT observations were combined on a phase-by-phase basis using \texttt{xselect} to maximize signal-to-noise, and grouped to a minimum of 20 counts per bin. For \textit{AstroSat} LAXPC, phase-resolved spectra from the two $\rho$-class observations were likewise combined per phase and grouped to a minimum of 30 counts per bin.

Power density spectra were computed with fractional-rms normalization using the \texttt{Stingray} package \citep{stingray} over 1024\,s segments. Light curves were binned at 1\,s for \textit{Swift} XRT and 0.01\,s for LAXPC. Energy-dependent time lags were derived between a broad reference band (0.3--10\,keV for XRT; 3--30\,keV for LAXPC) and narrow energy bins, with uncertainties estimated via bootstrap resampling. Dead-time corrections were applied to LAXPC data prior to any timing computation.

\section{Results}
\label{sec:results}

\subsection{Detection and Characterization of \texorpdfstring{$\rho$}{rho}-Class Oscillations}

Power density spectra computed from the phase-resolved light curves show clear evidence of quasi-periodic oscillations. The Leahy-normalized power density spectra (shown alongside the light curves in Figure~\ref{fig:rho_examples}) exhibit prominent peaks at $\sim$0.01--0.02\,Hz, corresponding to the characteristic $\sim$50--100\,s period of the $\rho$-class oscillations. The detection of this coherent periodicity in multiple observations from both \textit{Swift} XRT and \textit{AstroSat} LAXPC confirms the presence of $\rho$-class behavior and supports our phase-segmentation approach.

\begin{figure}[htbp]
    \centering
    \includegraphics[width=1\linewidth]{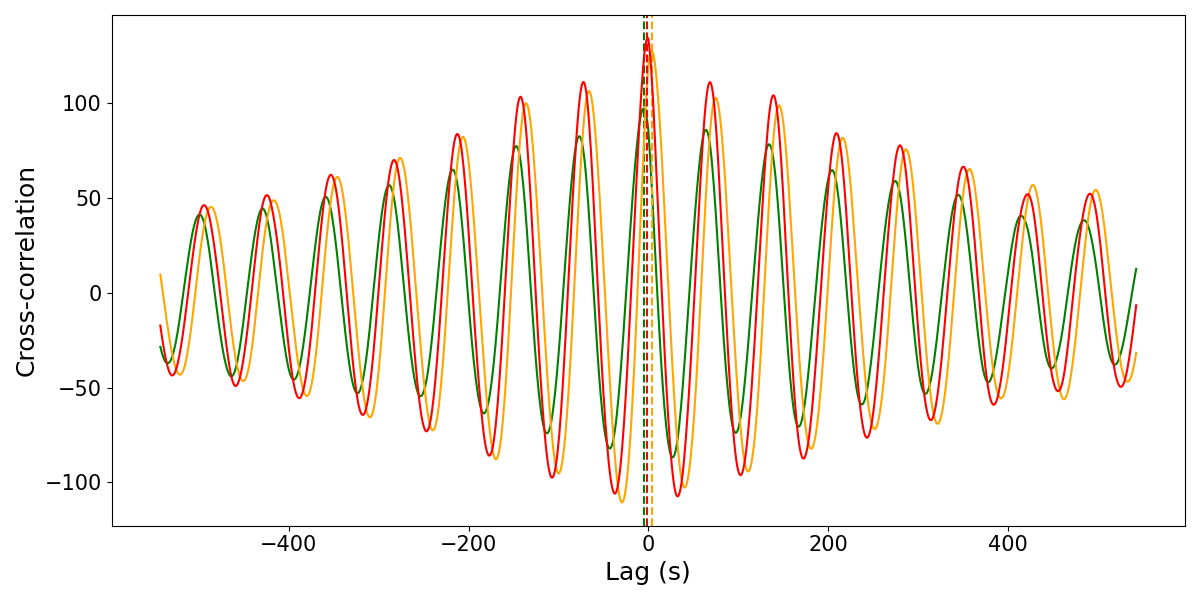}
    \caption{Cross-correlation functions between the three \textit{Swift} XRT energy bands of the $\rho$-class oscillations (OBSID 00030333112), computed using Savitzky--Golay smoothed light curves. The pairs low (0.3--1\,keV) vs mid (1--3\,keV) in green, mid (1--3\,keV) vs high (3--10\,keV) in yellow, and low (0.3--1\,keV) vs high (3--10\,keV) in red are shown. The low-energy band lags the mid band by $\sim$5\,s, while the mid band lags the high band by $\sim$4\,s, consistent with propagation of the variability from higher to lower energies. See Section~\ref{sec:results}.}
    \label{fig:cross_correlation}
\end{figure}

The cross-correlation analysis presented in Figure~\ref{fig:cross_correlation} shows systematic time delays between energy bands, with the high-energy variability leading the softer bands by several seconds. This energy-dependent behavior is consistent with the inward propagation of accretion rate fluctuations \citep{Uttley2014} and the outward propagation of the resulting disk structural changes from the inner to the outer disk regions.

\subsection{Phase-Resolved Spectral Evolution}

We fitted the phase-combined \textit{Swift} XRT spectra (1--10\,keV) using the model \texttt{tbabs*(diskbb + bremss + powerlaw)} in XSPEC\footnote{\url{https://heasarc.gsfc.nasa.gov/docs/xanadu/xspec/}}, where \texttt{tbabs} accounts for interstellar absorption, \texttt{diskbb} represents the multi-temperature thermal disk emission \citep{Mitsuda1984}, \texttt{bremss} models thermal bremsstrahlung from a hot corona \citep{Kellogg1975}, and \texttt{powerlaw} accounts for a steep hard X-ray tail. The neutral hydrogen column density $N_{\rm H}$ was left as a free parameter but remained consistent with Galactic values ($N_{\rm H} \sim 7$--$9 \times 10^{22}$\,cm$^{-2}$) across all phases. The power-law photon index was fixed at $\Gamma = -3$, as this component contributes negligibly to the total flux in the XRT band and consistently converged to this value across all phases.

\begin{figure*}[tbp]
    \centering
    \includegraphics[width=0.451\linewidth, angle=-90]{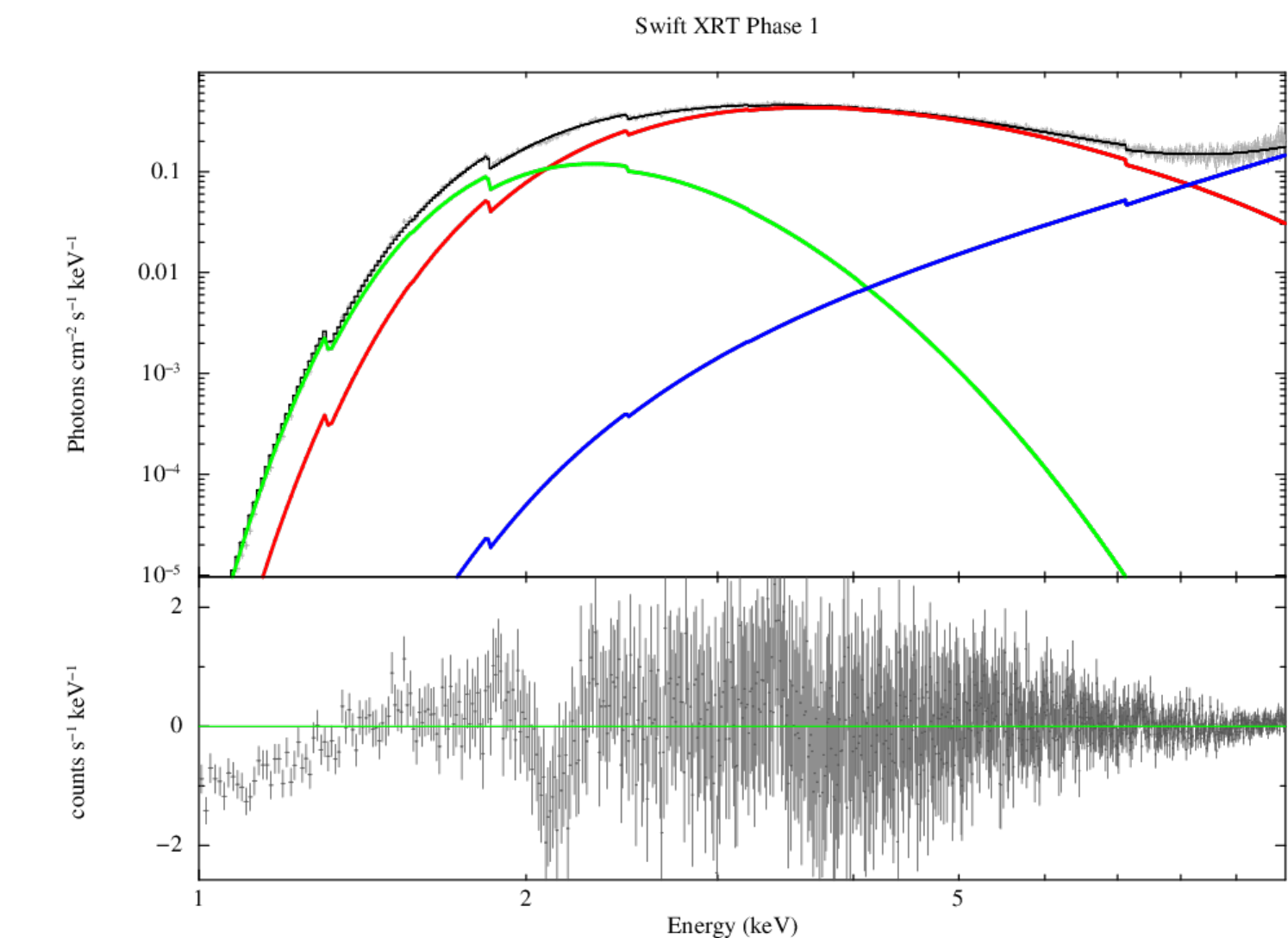}\\[0.5em]
    \includegraphics[width=0.62\linewidth]{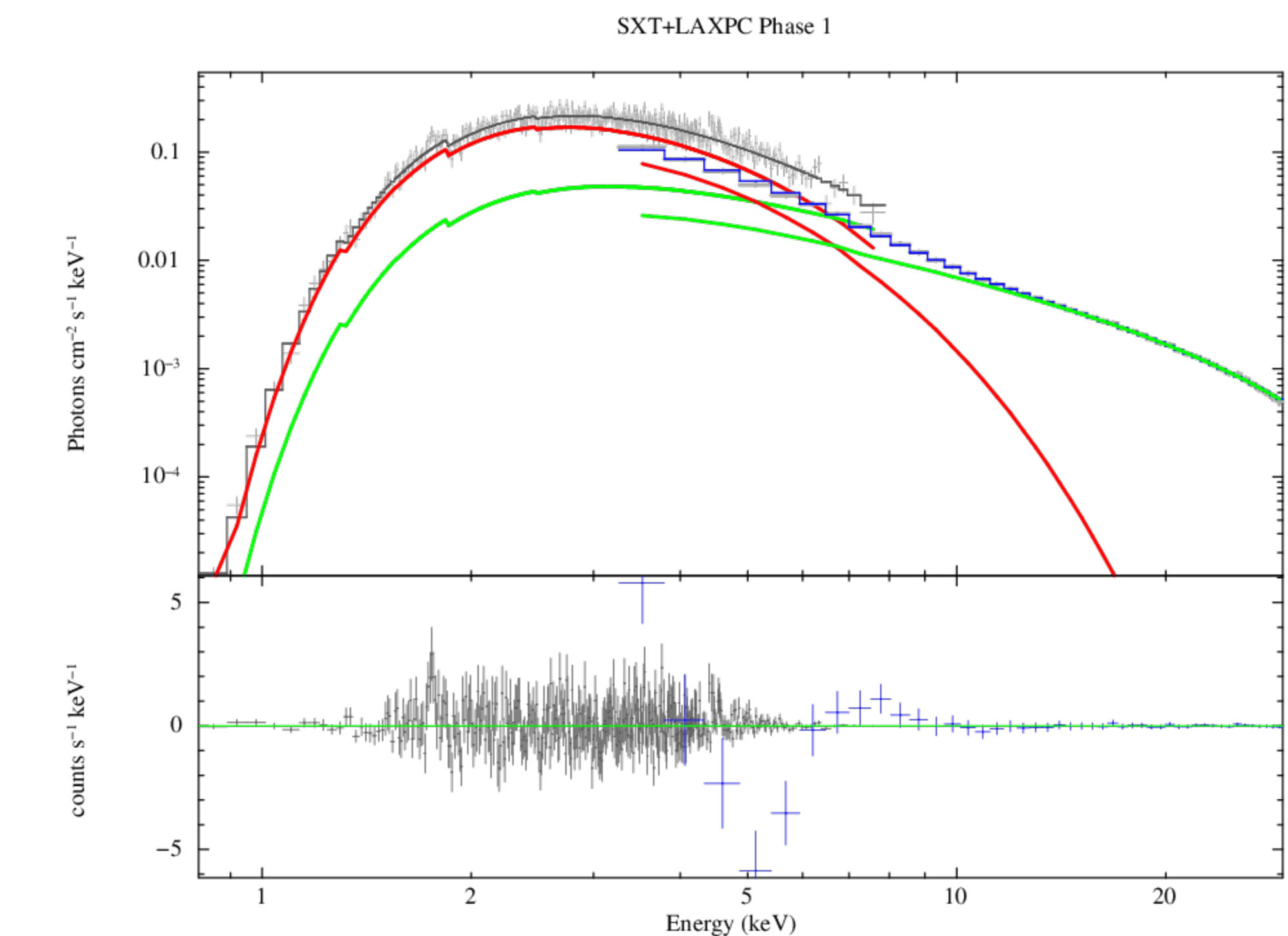}
    \caption{\textit{Top:} Unfolded Swift XRT spectrum for Phase~1 (rise I) of the $\rho$-cycle (1--10\,keV). Data points are shown in light gray, and the total best-fit model is plotted in black. The individual spectral components are overlaid: thermal bremsstrahlung (bremss) in green, multicolor disk blackbody (diskbb) in red, and power-law continuum in blue. \textit{Bottom:} Unfolded broadband \textit{AstroSat} SXT+LAXPC spectrum for Phase~1 of the $\rho$-cycle (0.8--30\,keV). The total best-fit model (\texttt{tbabs*const*(diskbb+nthcomp)}) is shown in black, with the thermal disk component (diskbb) in red and the Comptonized continuum (nthcomp) in green. Lower sub-panels show the fit residuals for each instrument.}
    \label{fig:phase1_spectrum}
    \label{fig:astrosat_spectrum}
\end{figure*}

The best-fit parameters for each phase are presented in Table~\ref{tab:xrt_params}. A representative unfolded spectrum for Phase~1 is shown in Figure~\ref{fig:phase1_spectrum}, illustrating the dominance of the thermal disk component (red) with a modest contribution from thermal bremsstrahlung (green). All fits yielded acceptable reduced $\chi^2$ values ranging from 1.09 to 1.20, confirming the statistical adequacy of the model.

\begin{table*}[tbp]
    \renewcommand{\arraystretch}{1.0}
    \caption{Phase-Resolved Spectral Fit Parameters at 90\% Confidence. \textit{Top:} \textit{Swift} XRT (1--10\,keV). \textit{Bottom:} \textit{AstroSat} SXT+LAXPC (0.8--30\,keV). $^{\rm f}$ denotes fixed parameter.}
    \begin{center}
    \footnotesize
    \begin{tabular}{llccccc}
        \toprule
        \textbf{Component} & \textbf{Parameter (Units)} & \textbf{Phase 1} & \textbf{Phase 2} & \textbf{Phase 3} & \textbf{Phase 4} & \textbf{Phase 5} \\ 
        \midrule
        \multicolumn{7}{c}{\textit{Swift} XRT: \texttt{tbabs*(diskbb + bremss + powerlaw)}} \\
        \midrule
        \textbf{TBabs} & $N_{\rm H}$ ($10^{22}$ cm$^{-2}$)
        & $7.20_{-0.22}^{+0.21}$ & $7.33_{-0.22}^{+0.21}$ & $8.62_{-0.19}^{+0.19}$ & $8.52_{-0.22}^{+0.21}$ & $7.36_{-0.20}^{+0.31}$ \\ 
        \midrule
        \multirow{2}{*}{\textbf{Diskbb}}
        & $T_{\rm in}$ (keV)
        & $1.69_{-0.03}^{+0.03}$ & $1.59_{-0.02}^{+0.03}$ & $1.53_{-0.02}^{+0.02}$ & $1.91_{-0.03}^{+0.04}$ & $1.89_{-0.06}^{+0.03}$ \\  
        & $R_{\rm in}$ (km)
        & $22.4_{-1.0}^{+1.0}$ & $28.3_{-1.1}^{+1.2}$ & $37.6_{-1.3}^{+1.4}$ & $23.4_{-1.0}^{+1.1}$ & $18.1_{-0.7}^{+1.4}$ \\  
        \midrule
        \multirow{2}{*}{\textbf{Bremss}}
        & $kT$ (keV)
        & $0.485_{-0.022}^{+0.026}$ & $0.457_{-0.019}^{+0.022}$ & $0.363_{-0.010}^{+0.010}$ & $0.375_{-0.011}^{+0.012}$ & $0.458_{-0.027}^{+0.019}$ \\  
        & Norm
        & $224_{-65}^{+83}$ & $342_{-97}^{+120}$ & $2530_{-540}^{+680}$ & $1780_{-410}^{+530}$ & $281_{-62}^{+140}$ \\  
        \midrule
        \multirow{2}{*}{\textbf{Powerlaw}}
        & $\Gamma^{\rm f}$
        & $-3$ & $-3$ & $-3$ & $-3$ & $-3$ \\  
        & Norm
        & $0.00016$ & $0.00024$ & $0.00049$ & $0.00054$ & $0.00019$ \\  
        \midrule
        \multirow{3}{*}{\textbf{Fit}}
        & Flux ($10^{-8}$ erg\,cm$^{-2}$\,s$^{-1}$)
        & $1.8$ & $2.3$ & $3.5$ & $3.9$ & $2.0$ \\  
        & $\chi^2$
        & $972$ & $986$ & $1067$ & $1016$ & $1029$ \\  
        & $\chi^2_\nu$
        & $1.09$ & $1.11$ & $1.20$ & $1.14$ & $1.15$ \\ 
        \midrule
        \midrule
        \multicolumn{7}{c}{\textit{AstroSat} SXT+LAXPC: \texttt{tbabs*const*(diskbb + nthcomp)}} \\
        \midrule
        \textbf{TBabs} & $N_{\rm H}$ ($10^{22}$ cm$^{-2}$)
        & $3.37_{-0.08}^{+0.13}$ & $3.46_{-0.09}^{+0.11}$ & $3.73_{-0.10}^{+0.10}$ & $3.85_{-0.12}^{+0.15}$ & $3.56_{-0.13}^{+0.16}$ \\
        \midrule
        \multirow{2}{*}{\textbf{Constant}}
        & SXT$^{\rm f}$
        & $1.0$ & $1.0$ & $1.0$ & $1.0$ & $1.0$ \\
        & LAXPC
        & $0.55_{-0.02}^{+0.01}$ & $0.65_{-0.02}^{+0.02}$ & $0.63_{-0.02}^{+0.02}$ & $0.61_{-0.02}^{+0.02}$ & $0.67_{-0.02}^{+0.02}$ \\
        \midrule
        \multirow{2}{*}{\textbf{Diskbb}}
        & $T_{\rm in}$ (keV)
        & $1.31_{-0.05}^{+0.05}$ & $1.26_{-0.05}^{+0.04}$ & $1.27_{-0.04}^{+0.04}$ & $1.24_{-0.05}^{+0.04}$ & $1.31_{-0.06}^{+0.05}$ \\
        & $R_{\rm in}$ (km)
        & $17.7_{-0.9}^{+1.4}$ & $19.8_{-1.1}^{+1.4}$ & $21.8_{-1.2}^{+1.4}$ & $22.9_{-1.4}^{+1.8}$ & $17.4_{-1.1}^{+1.4}$ \\
        \midrule
        \multirow{3}{*}{\textbf{Nthcomp}}
        & $\Gamma$
        & $1.77_{-0.10}^{+0.10}$ & $1.94_{-0.08}^{+0.08}$ & $1.93_{-0.08}^{+0.08}$ & $2.01_{-0.08}^{+0.09}$ & $2.00_{-0.11}^{+0.12}$ \\
        & $kT_{\rm e}$ (keV)
        & $6.2_{-0.5}^{+0.6}$ & $10.2_{-1.4}^{+2.5}$ & $10.5_{-1.5}^{+2.7}$ & $14.5_{-3.4}^{+10.3}$ & $11.5_{-2.5}^{+7.0}$ \\
        & Norm
        & $0.23_{-0.05}^{+0.06}$ & $0.38_{-0.06}^{+0.07}$ & $0.41_{-0.07}^{+0.07}$ & $0.51_{-0.08}^{+0.09}$ & $0.36_{-0.07}^{+0.09}$ \\
        \midrule
        \multirow{2}{*}{\textbf{Fit}}
        & $\chi^2$
        & $379$ & $439$ & $454$ & $379$ & $232$ \\
        & $\chi^2_\nu$
        & $0.93$ & $1.04$ & $1.08$ & $1.10$ & $0.84$ \\
        \bottomrule
    \end{tabular}
    \end{center}
    \label{tab:xrt_params}
    \label{tab:astrosat_params}
\end{table*}

\subsubsection{Inner Disk Temperature and Radius Evolution from \textit{Swift} XRT}

\new{Taken at face value, the \textit{Swift} XRT fits show} a systematic anti-correlation between the inner disk temperature $T_{\rm in}$ and the apparent inner disk radius $R_{\rm in}$ throughout the $\rho$ cycle. As listed in the XRT block of Table~\ref{tab:xrt_params}, $T_{\rm in}$ decreases from $\sim$1.69\,keV in Phase~1 through $\sim$1.59\,keV in Phase~2 to a minimum of $\sim$1.53\,keV in Phase~3, while $R_{\rm in}$ increases from $\sim$22\,km in Phase~1 to a maximum of $\sim$37.6\,km in Phase~3. At Phase~4 (the burst peak), the apparent disk collapses inward ($R_{\rm in} \sim 23.4$\,km) and heats ($T_{\rm in} \sim 1.91$\,keV). In Phase~5, the post-burst disk contracts further ($R_{\rm in} \sim 18.1$\,km, $T_{\rm in} \sim 1.89$\,keV). \new{We show in Section~\ref{sec:tin_constancy} that the broadband \textit{AstroSat} fits do not statistically require this disk-temperature variation, and we interpret the larger XRT swings as a consequence of the limited 1--10\,keV bandpass (Section~\ref{sec:rin_discrepancy}).}

The total 1--10\,keV flux remains nearly constant between Phase~3 ($3.5 \times 10^{-8}$\,erg\,cm$^{-2}$\,s$^{-1}$) and Phase~4 ($3.9 \times 10^{-8}$\,erg\,cm$^{-2}$\,s$^{-1}$) despite a $\sim$25\% increase in $T_{\rm in}$, because the decrease in $R_{\rm in}^2$ largely compensates for the increase in $T_{\rm in}^4$. This near-constant luminosity during the collapse is consistent with radiation-pressure driving, where the total energy output is set by the accretion rate rather than the disk geometry at the time \citep{frank2002accretion, abramowicz1988slim}.

\subsection{Broadband Spectral Analysis with \textit{AstroSat}}

To overcome the limited energy range of \textit{Swift} XRT and to constrain the Comptonized continuum component, we performed joint spectral fitting of \textit{AstroSat} SXT (0.8--7\,keV) and LAXPC (3--30\,keV) data using the model \texttt{tbabs*const*(diskbb + nthcomp)}. Here, \texttt{const} accounts for cross-normalization between instruments, \texttt{diskbb} represents the thermal disk, and \texttt{nthcomp} models the Comptonized continuum from a hot corona \citep{Zdziarski1996}. The broader energy coverage of \textit{AstroSat} eliminated the need for an additional bremsstrahlung component, as the Comptonization accounts for the hard X-ray emission (see Section~\ref{sec:bremss} for discussion of the relationship between the XRT and \textit{AstroSat} spectral models).

The best-fit parameters are presented in Table~\ref{tab:astrosat_params}. The broadband \textit{AstroSat} fits show qualitatively similar trends to those observed in \textit{Swift} XRT, although with much smaller amplitudes: $T_{\rm in}$ varies only weakly across the cycle, from $\sim$1.31\,keV in Phase~1 to a minimum of $\sim$1.24\,keV at the burst peak (Phase~4), while $R_{\rm in}$ increases from $\sim$17.7\,km in Phase~1 to a maximum of $\sim$22.9\,km at the burst peak (Phase~4) and contracts back to $\sim$17.4\,km in Phase~5. The absolute values of $R_{\rm in}$ from \textit{AstroSat} are systematically smaller than those from \textit{Swift} XRT (17--23\,km vs.\ 18--38\,km), which we attribute to the different spectral models and energy ranges employed; this is discussed further in Section~\ref{sec:rin_discrepancy}. We return to the question of whether the apparent disk-temperature variation in the \textit{AstroSat} fits is statistically required in Section~\ref{sec:tin_constancy}.

Figure~\ref{fig:astrosat_spectrum} shows the unfolded broadband spectrum for Phase~1, with the thermal disk component (red) dominating below $\sim$7\,keV and the Comptonized component (green) extending to higher energies. The good fit quality (reduced $\chi^2 \sim 0.84$--1.10) indicates that this two-component model adequately describes the continuum across 0.8--30\,keV. \new{The narrow per-phase span of $T_{\rm in}$ in the \textit{AstroSat} fits prompts a direct test of whether the data require any temperature variation through the cycle.}

\subsection{\new{Joint Test of Inner Disk Temperature Constancy}}
\label{sec:tin_constancy}

{
To assess whether the apparent $T_{\rm in}$ variation in Table~\ref{tab:astrosat_params} is statistically required by the broadband data, we performed a joint analysis of all five phases simultaneously (10 datasets: five SXT and five LAXPC) in three configurations: (i) all parameters free per phase, which reproduces the per-phase fit summarized in Table~\ref{tab:astrosat_params}; (ii) $T_{\rm in}$ tied to a single global value across the cycle while every other parameter is allowed to vary per phase; and (iii) both $T_{\rm in}$ and the \texttt{diskbb} normalization tied across the cycle. The SXT--LAXPC cross-normalization was kept free per phase in all three configurations to preserve the per-observation flux offsets between instruments.

Tying $T_{\rm in}$ across all five phases costs only $\Delta\chi^2 = +4.3$ for 4 added degrees of freedom, which is fully consistent with statistical fluctuation; the joint reduced chi-squared remains $\chi^2_\nu = 1.003$ for 1881 dof, statistically identical to the per-phase fit. The best-fit global temperature is $T_{\rm in} = 1.275 \pm 0.020$\,keV (90\% CI). The broadband \textit{AstroSat} data are therefore statistically consistent with a single constant inner disk temperature throughout the $\rho$ cycle.

Tying the disk normalization in addition to $T_{\rm in}$, however, is rejected at high significance: $\Delta\chi^2 = +175.9$ for 8 added degrees of freedom relative to the per-phase fit. Under this combined constraint the corona is forced to absorb the entire residual phase-to-phase variation, with $kT_{\rm e}$ running off to unphysical values ($\gtrsim 10^3$\,keV in Phases~3--5) and the fit becoming statistically poor. This rules out a fully constant disk and confirms that genuine variation in the apparent disk emitting area is required by the data.

We checked that this conclusion is robust to the details of the analysis. An $F$-test comparing the tied- and free-$T_{\rm in}$ configurations returns a probability well above $0.05$ for the four additional degrees of freedom, so the free-temperature model is not statistically preferred at any meaningful confidence level. Repeating the joint fit with the SXT--LAXPC cross-normalization tied to a single value across phases, rather than left free, shifts the global temperature by less than its $90\%$ statistical uncertainty and leaves the $\Delta\chi^2$ for tying $T_{\rm in}$ below the threshold for significance. We also confirmed that the result is insensitive to the choice of energy binning and to the lower-energy bound of the SXT band within the calibrated range. The constancy of the inner disk temperature is therefore not an artifact of the cross-calibration treatment or of the binning, but a genuine property of the broadband data.
}

\subsection{\new{Phase-Resolved Parameters from the \texorpdfstring{$T_{\rm in}$}{Tin}-Tied Joint Fit}}
\label{sec:joint_phase_params}

{
The phase-resolved coronal parameters from the $T_{\rm in}$-tied fit (Table~\ref{tab:joint_constancy}) follow a monotonic evolution across the cycle. The electron temperature $kT_{\rm e}$ rises from $\sim$6.5\,keV in Phase~1 to $\sim$13.8\,keV in Phase~5, and the photon index $\Gamma$ steepens from $\sim$1.82 in Phase~1 to $\sim$2.06 in Phase~5. The cross-normalization constants float between $0.56$ and $0.68$ across phases. The inner-disk emitting area, freed from the $T_{\rm in}$ constraint via the disk normalization, varies modestly through the cycle.

\begin{table*}[tbp]
    \renewcommand{\arraystretch}{1.0}
    \caption{\new{Joint fit of all five \textit{AstroSat} phases (10 datasets) with $T_{\rm in}$ tied to a single global value across the cycle while every other parameter is free per phase. Cross-normalization is fixed at SXT~=~1.0 with LAXPC free per phase. Errors are 90\% confidence intervals. The bottom rows summarize the joint fit statistics and the comparison with the (rejected) configuration in which both $T_{\rm in}$ and the disk normalization are tied.} \label{tab:joint_constancy}}
    \begin{center}
    \footnotesize
    \begin{tabular}{llccccc}
        \toprule
        \textbf{Component} & \textbf{Parameter (Units)} & \textbf{Phase 1} & \textbf{Phase 2} & \textbf{Phase 3} & \textbf{Phase 4} & \textbf{Phase 5} \\
        \midrule
        \textbf{TBabs} & $N_{\rm H}$ ($10^{22}$ cm$^{-2}$)
        & $3.41_{-0.06}^{+0.07}$ & $3.44_{-0.06}^{+0.06}$ & $3.72_{-0.07}^{+0.07}$ & $3.81_{-0.08}^{+0.09}$ & $3.61_{-0.09}^{+0.09}$ \\
        \midrule
        \multirow{2}{*}{\textbf{Constant}}
        & SXT$^{\rm f}$
        & $1.0$ & $1.0$ & $1.0$ & $1.0$ & $1.0$ \\
        & LAXPC
        & $0.56_{-0.01}^{+0.01}$ & $0.65_{-0.01}^{+0.02}$ & $0.63_{-0.01}^{+0.01}$ & $0.61_{-0.02}^{+0.02}$ & $0.68_{-0.02}^{+0.02}$ \\
        \midrule
        \multirow{2}{*}{\textbf{Diskbb}}
        & $T_{\rm in}$ (keV) (tied)
        & \multicolumn{5}{c}{$1.275_{-0.020}^{+0.020}$} \\
        & $R_{\rm in}$ (km)
        & $18.48_{-0.57}^{+0.62}$ & $19.47_{-0.61}^{+0.66}$ & $21.72_{-0.67}^{+0.73}$ & $21.94_{-0.72}^{+0.78}$ & $18.14_{-0.63}^{+0.68}$ \\
        \midrule
        \multirow{3}{*}{\textbf{Nthcomp}}
        & $\Gamma$
        & $1.82_{-0.06}^{+0.06}$ & $1.92_{-0.05}^{+0.05}$ & $1.92_{-0.06}^{+0.06}$ & $1.97_{-0.06}^{+0.06}$ & $2.06_{-0.07}^{+0.08}$ \\
        & $kT_{\rm e}$ (keV)
        & $6.5_{-0.4}^{+0.5}$ & $9.8_{-1.0}^{+1.5}$ & $10.4_{-1.2}^{+1.9}$ & $12.5_{-2.0}^{+3.8}$ & $13.8_{-3.1}^{+8.7}$ \\
        & Norm
        & $0.27_{-0.03}^{+0.04}$ & $0.37_{-0.04}^{+0.04}$ & $0.41_{-0.04}^{+0.05}$ & $0.46_{-0.05}^{+0.05}$ & $0.41_{-0.05}^{+0.05}$ \\
        \midrule
        \multicolumn{2}{l}{\textbf{Joint fit:}} & \multicolumn{5}{l}{$\chi^2 = 1887$, dof = 1881, $\chi^2_\nu = 1.003$} \\
        \multicolumn{2}{l}{$\Delta\chi^2$ vs.\ free per-phase fit:} & \multicolumn{5}{l}{$+4.3$ for $\Delta$dof = $+4$ (consistent with statistical fluctuation)} \\
        \multicolumn{2}{l}{$\Delta\chi^2$ when both $T_{\rm in}$ and disk norm tied:} & \multicolumn{5}{l}{$+175.9$ for $\Delta$dof = $+8$ (rejected)} \\
        \bottomrule
    \end{tabular}
    \end{center}
\end{table*}
Rescaling $R_{\rm in}$ from the joint disk normalizations (anchored to the per-phase Phase~4 value of $R_{\rm in} = 22.9$\,km at $\mathrm{norm}_{\rm disk} = 211$) gives $R_{\rm in} = 18.5 \pm 0.6$, $19.5 \pm 0.7$, $21.7 \pm 0.7$, $21.9 \pm 0.8$, and $18.1 \pm 0.7$\,km in Phases~1--5 respectively. The variation amplitude is $\sim$20\% with $1\sigma$-equivalent errors of $\sim$3--4\%, separating the Phase~1/5 minima and the Phase~4 maximum at well above the $4\sigma$ level.
}
\subsection{Hardness--Intensity and Color--Color Diagrams}

To investigate spectral state transitions during the $\rho$ cycle, we constructed Hardness--Intensity Diagrams (HIDs) and Color--Color Diagrams (CCDs) for each instrument. For \textit{Swift} XRT, intensity was defined as the total 1.0--10.0\,keV count rate, and hardness as the ratio of counts in the 5.5--10.0\,keV band to the 1.0--5.5\,keV band. Soft and hard colors were defined as the ratios of counts in the 3.0--4.5\,keV to 1.0--3.0\,keV and 4.5--7.0\,keV to 1.0--3.0\,keV bands, respectively. Similar definitions were adopted for \textit{AstroSat} SXT and LAXPC, adjusted for their respective energy ranges.

In each diagram the five phase bins are color-coded following the scheme of Figure~\ref{fig:phase_classification} (Phases~1--3 in red, green, and blue; Phases~4--5 in cyan and gray), and every plotted point corresponds to a one-second time bin assigned to its phase by the segmentation procedure of Section~\ref{sec:phase}. Plotting all phases jointly allows the trajectory of the source through the diagram to be followed directly, while the spread of points within a given phase reflects the cycle-to-cycle scatter of the heartbeat oscillation. We constructed the diagrams separately for each instrument so that the soft-band behavior seen by \textit{Swift} XRT and \textit{AstroSat} SXT could be compared against the hard-band behavior seen by LAXPC over the same set of phases.

\begin{figure*}[tbp]
    \centering
    \includegraphics[width=0.48\textwidth]{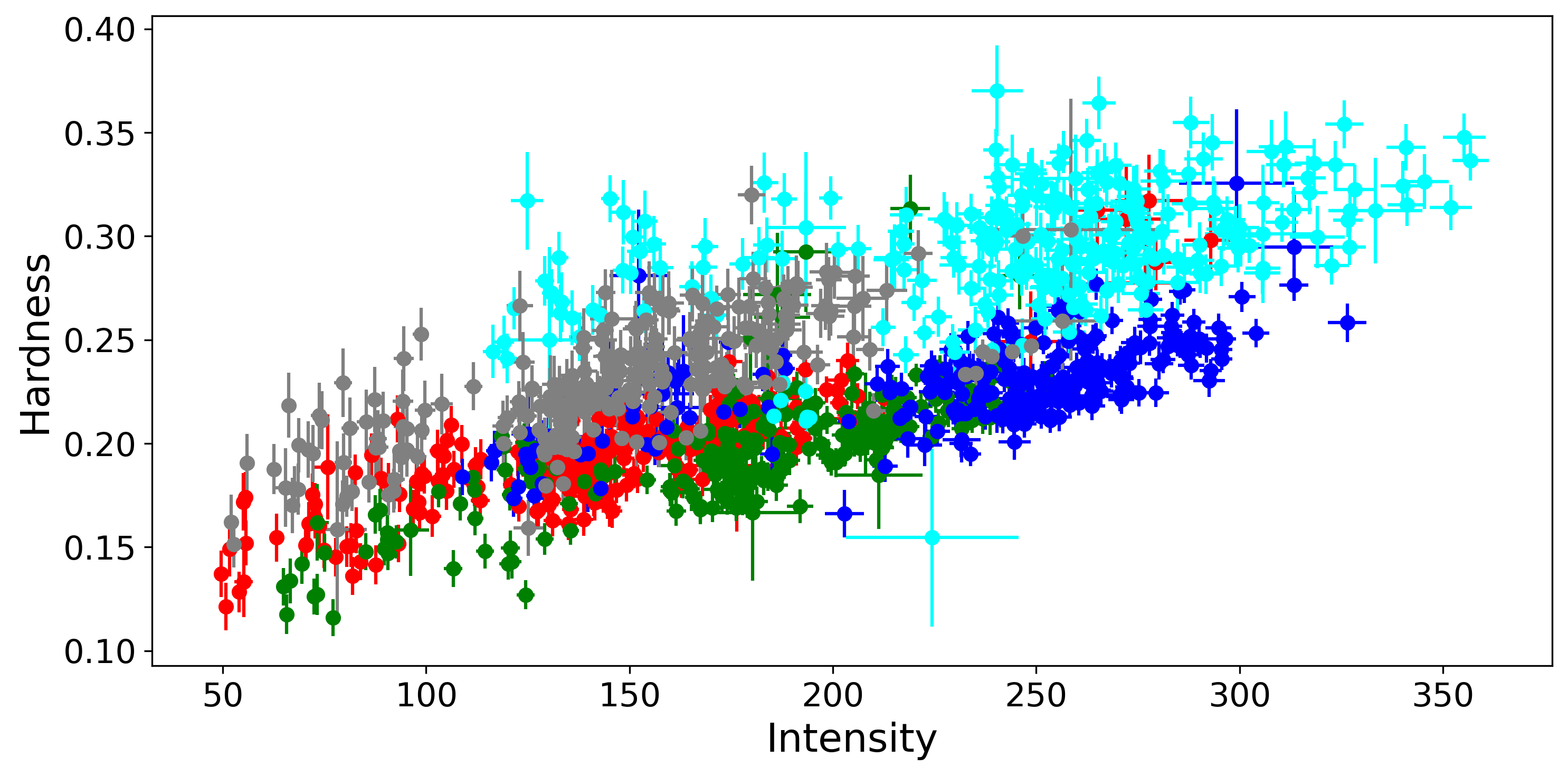}
    \hfill
    \includegraphics[width=0.48\textwidth]{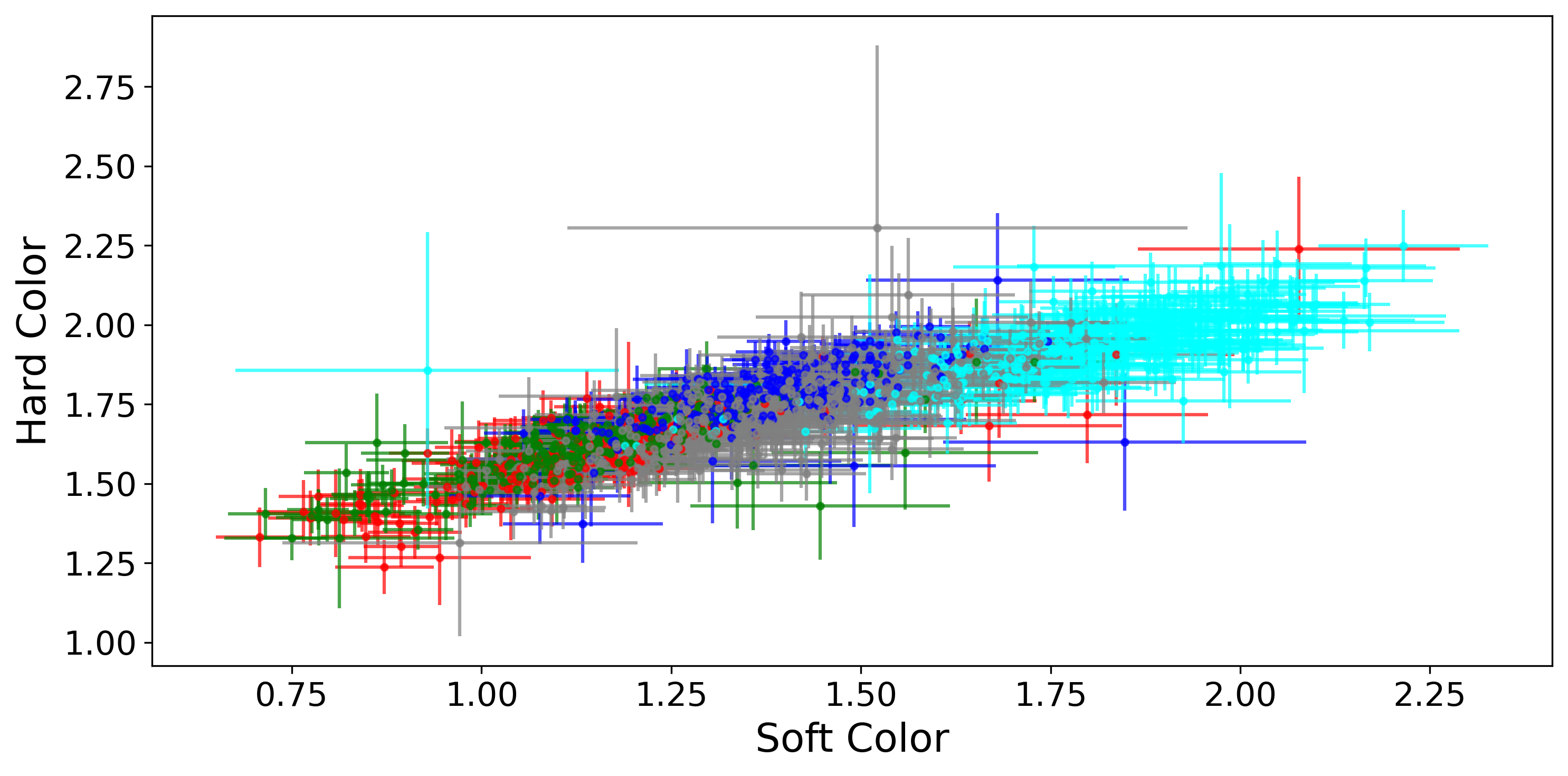}\\[0.5em]
    \includegraphics[width=0.48\textwidth]{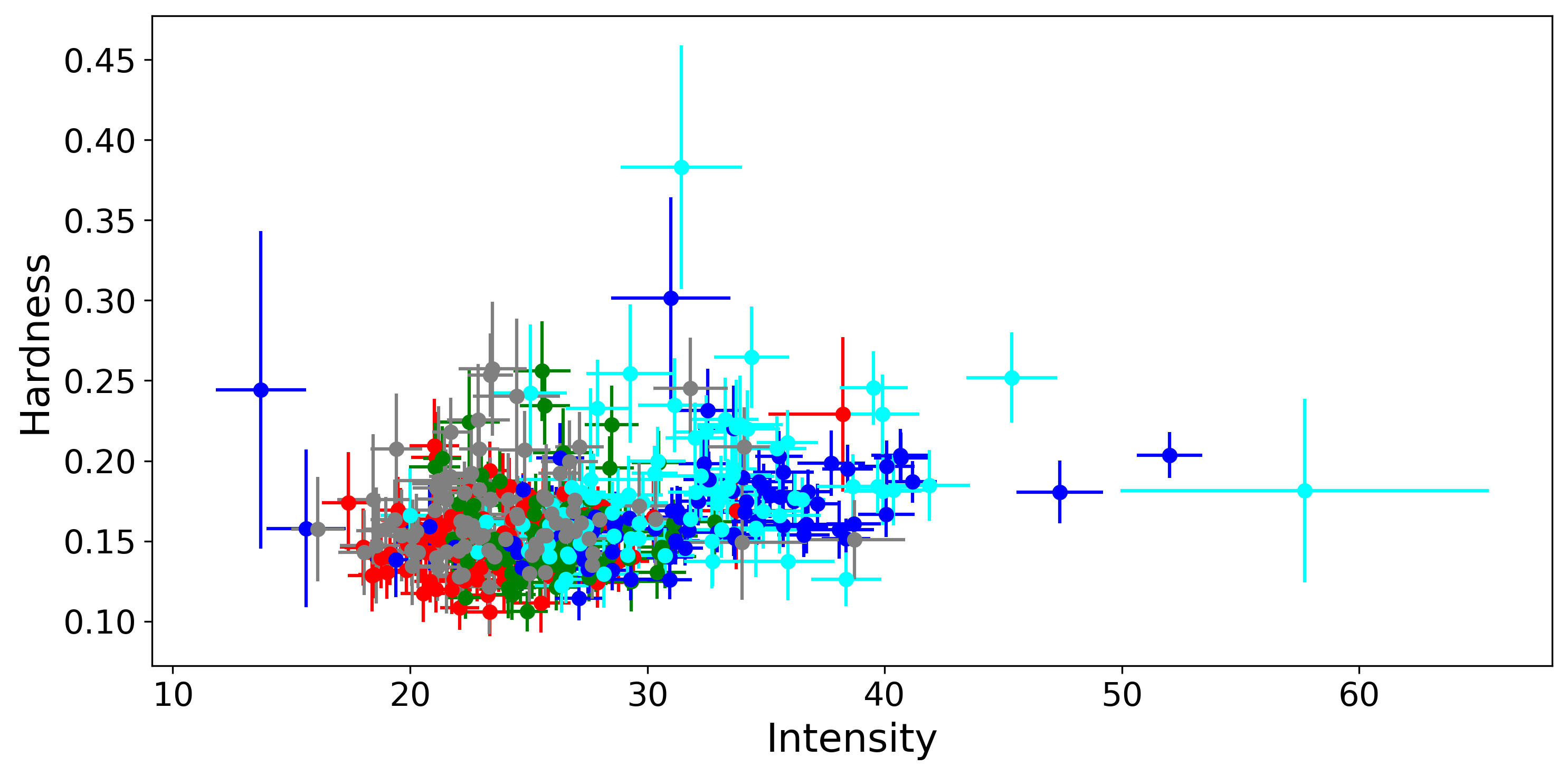}
    \hfill
    \includegraphics[width=0.48\textwidth]{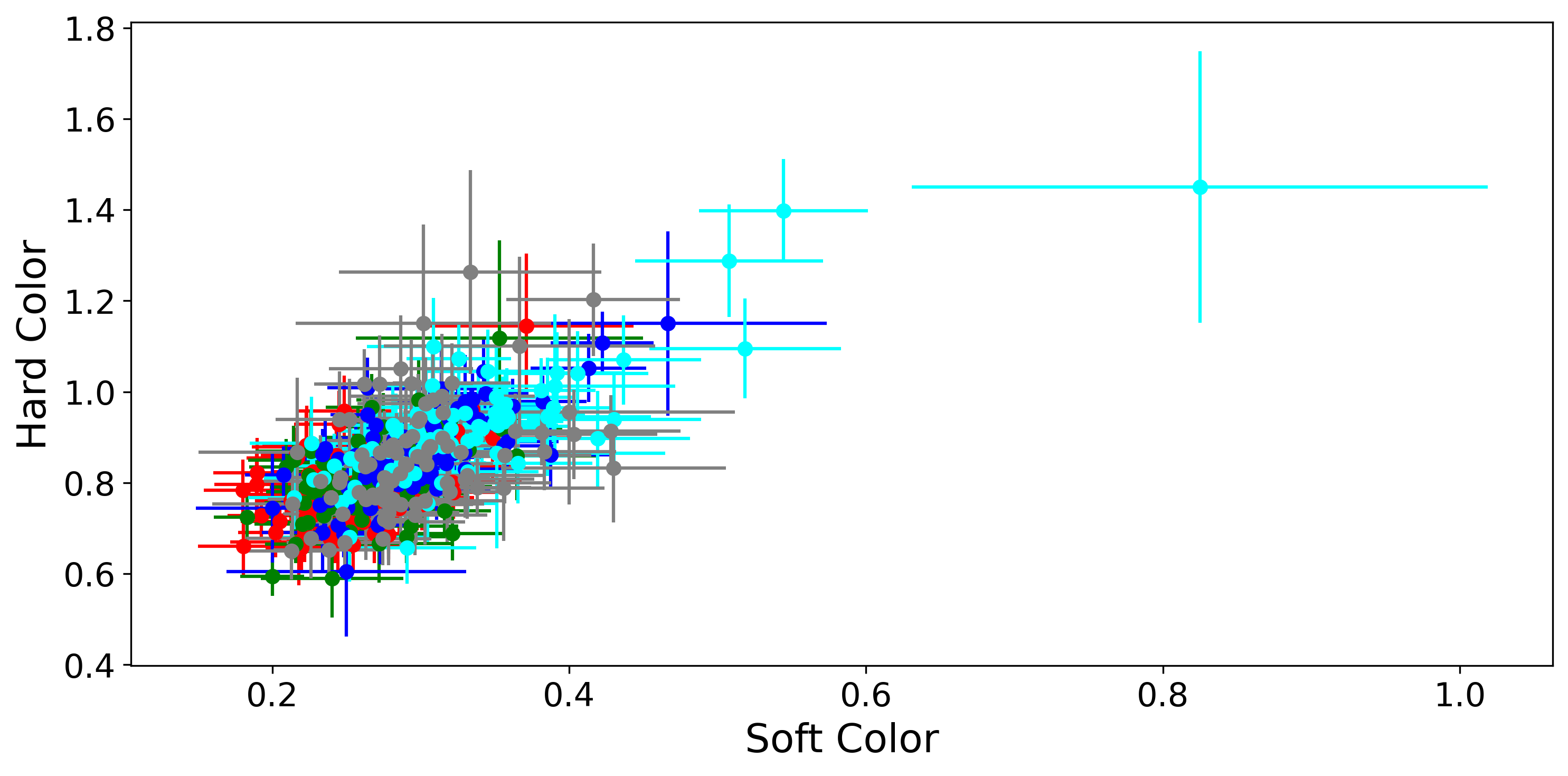}\\[0.5em]
    \includegraphics[width=0.48\textwidth]{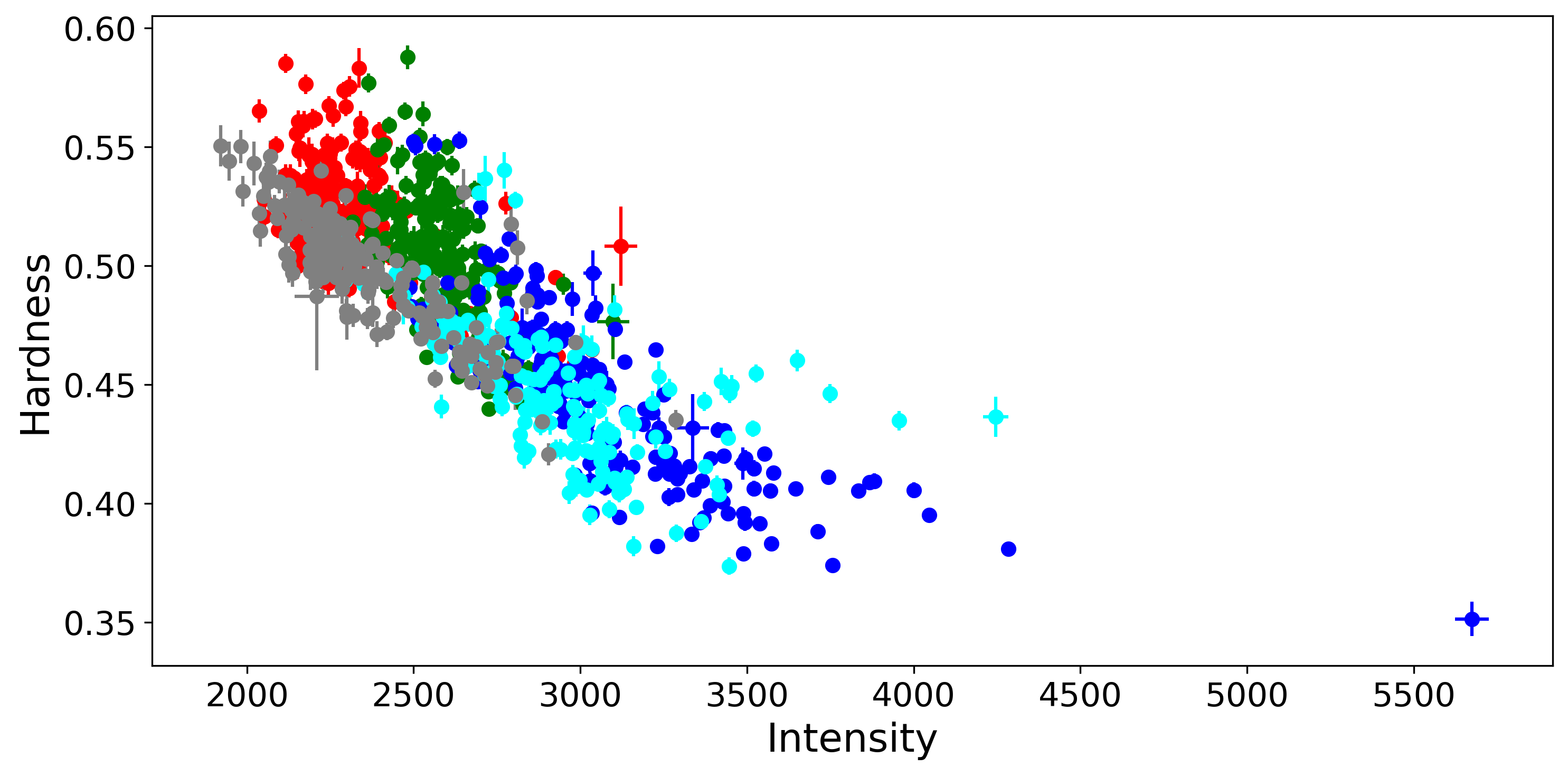}
    \hfill
    \includegraphics[width=0.48\textwidth]{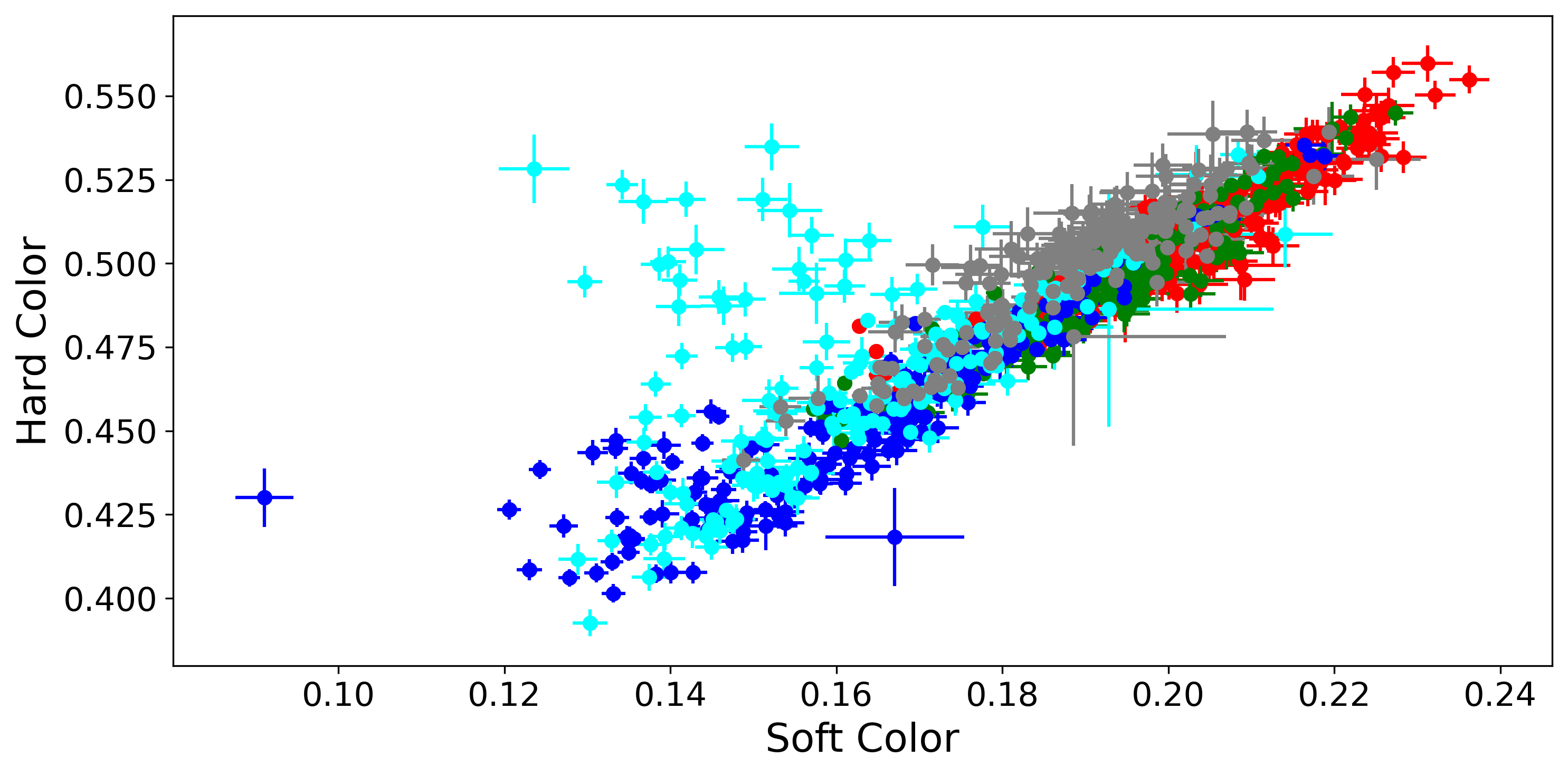}
    \caption{Hardness--Intensity Diagrams (HIDs, left column) and Color--Color Diagrams (CCDs, right column) for all three instruments across the five $\rho$-cycle phases (colors as in Figure~\ref{fig:phase_classification}). \textit{Top row, Swift XRT:} HID intensity = 1.0--10.0\,keV count rate, hardness = 5.5--10.0\,keV / 1.0--5.5\,keV; CCD soft color = 3.0--4.5\,keV / 1.0--3.0\,keV, hard color = 4.5--7.0\,keV / 1.0--3.0\,keV. \textit{Middle row, AstroSat SXT:} HID intensity = 1.0--7.0\,keV count rate, hardness = 4.5--7.0\,keV / 1.0--4.5\,keV; CCD definitions identical to XRT. \textit{Bottom row, AstroSat LAXPC:} HID intensity = 3.0--30.0\,keV count rate, hardness = 6.0--30.0\,keV / 3.0--6.0\,keV; CCD soft color = 7.0--15.0\,keV / 3.0--7.0\,keV, hard color = 15.0--30.0\,keV / 3.0--7.0\,keV. Each phase occupies a distinct region, with the system tracing a counter-clockwise hysteresis loop through the $\rho$ cycle. Phase~3 is the softest in XRT/SXT, while in LAXPC above $\sim$10\,keV the hardest spectrum is that of the low-intensity Phase~1, so the hardness ordering of the phases is not preserved across energy.}
    \label{fig:hid_xrt}
    \label{fig:ccd_xrt}
    \label{fig:hid_sxt}
    \label{fig:ccd_sxt}
    \label{fig:hid_laxpc}
    \label{fig:ccd_laxpc}
\end{figure*}

The resulting HIDs and CCDs show hysteresis patterns, most clearly in the \textit{Swift} XRT diagrams (Figure~\ref{fig:hid_xrt}). Each phase occupies a distinct region in these diagrams, with the system tracing a counter-clockwise loop through the $\rho$ cycle, reflecting the expansion-collapse sequence: the system moves to softer states during Phases~1--3 as the disk expands and cools, then to harder states at the burst (Phase~4) and post-burst contraction (Phase~5). Phase~3 corresponds to the softest state in \textit{Swift} XRT, consistent with the coolest disk temperature and largest radius in the XRT spectral fits, and similarly appears among the softest states in \textit{AstroSat} SXT where the disk has substantially expanded; Phase~4 occupies the highest-intensity region, corresponding to the burst peak.

The hysteresis is most pronounced in the \textit{Swift} XRT HID, where the rise phases (1--3) trace the lower, softer branch of the loop and the burst and decay phases (4--5) trace the upper, harder branch, so that the system does not retrace its path but encloses a finite area in the hardness--intensity plane. The \textit{AstroSat} SXT HID, restricted to a narrower soft band, shows a much more compact and overlapping distribution with no well-defined loop, consistent with the SXT sampling primarily the thermal disk whose temperature varies only weakly across the cycle. In the color--color diagrams, the soft and hard colors are positively correlated during the rise and decouple near the burst, producing the characteristic looping that distinguishes genuine spectral evolution from simple flux scaling.

The hardness ordering of the phases changes with energy band. Phase~3 is the softest phase in the soft X-ray bands (XRT/SXT), but it does not become the hardest at higher energies; in the LAXPC band the hardest spectrum is that of the low-intensity Phase~1, which carries the flattest Comptonized continuum ($\Gamma \sim 1.8$; Table~\ref{tab:astrosat_params}), while Phase~3 stays comparatively soft. This energy dependence indicates that thermal disk emission dominates the soft band while Comptonization governs the hardness ranking at higher energies \citep{done2007xrb}, which shows the need for broadband coverage to capture both components simultaneously.

\begin{figure*}[tbp]
    \centering
    \includegraphics[width=0.48\linewidth]{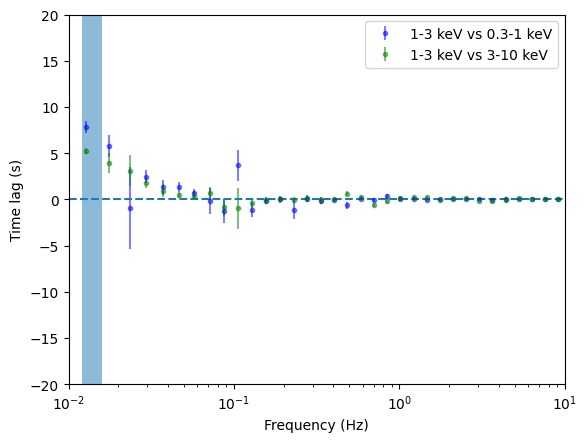}\hfill
    \includegraphics[width=0.48\linewidth]{wt\_po\_bary\_lag\_spectrum\_0.012-0.015_copy.png}
    \caption{\textit{Left:} Frequency-dependent time lags from \textit{Swift} XRT $\rho$-class data (OBSID 00030333112). Lags are shown between the 1--3\,keV reference band and the 0.3--1\,keV (blue) and 3--10\,keV (green) bands. Significant positive lags of 5--8\,s are detected at the heartbeat frequency ($\sim$0.01--0.02\,Hz, shaded), diminishing to near zero above $\sim$0.1\,Hz. \textit{Right:} Time-lag spectrum in the frequency range 0.012--0.015\,Hz. Positive (negative) lags indicate harder (softer) photons arriving later. The complex structure suggests contributions from multiple emission regions. See Section~\ref{sec:results} for discussion.}
    \label{fig:freq_dependent_lags}
    \label{fig:lag_spectrum}
\end{figure*}

The three instruments map complementary portions of the same spectral trajectory: the soft-band diagrams track the thermal disk while the LAXPC diagrams track the Comptonized corona, and the phase at which each instrument reaches its hardest state differs accordingly. The single-sense hysteresis seen most clearly in the \textit{Swift} XRT HID, together with the differing hardness orderings between the soft-band and LAXPC diagrams, confirms that the $\rho$ cycle is hysteretic and that the disk and corona respond to the underlying instability with a measurable phase offset. We quantify this offset directly through the energy-dependent time lags presented in Section~\ref{sec:results}.
\subsection{Timing Analysis: Energy-Dependent Time Lags}

Time-lag analysis was performed to investigate the propagation of variability between different energy bands. Time lags were computed between the reference and subject energy bands using Fourier cross-spectral techniques \citep{Uttley2014}. The reference band was taken as 1--3\,keV, and lags were measured relative to the 0.3--1\,keV and 3--10\,keV bands using a representative $\rho$-class observation (OBSID 00030333112).
The frequency-dependent time lags shown in Figure~\ref{fig:freq_dependent_lags} (left) exhibit positive lags at the heartbeat frequency ($\sim$0.01--0.02\,Hz), with both the soft and hard bands lagging behind the 1--3\,keV reference band by $\sim$5--8\,s. At frequencies above $\sim$0.1\,Hz, the lags diminish toward zero, consistent with the variability being confined to the heartbeat timescale. The lag spectrum (Figure~\ref{fig:freq_dependent_lags}, right) exhibits a structured pattern with both hard and soft lags at different frequencies, suggesting contributions from multiple physical processes such as disk reverberation, Comptonization time delays, and intrinsic propagation within the accretion flow \citep{Uttley2014}. The sign reversal between hard and soft lags at different frequencies indicates that these processes operate on distinct timescales within the $\rho$ cycle and are not confined to a single emission region.
\section{Discussion}
\label{sec:discussion}

\subsection{The radiation-pressure instability cycle}
\label{sec:radpressure}

\new{We first describe the disk evolution as inferred from the narrow-band \textit{Swift} XRT fits taken at face value; in Section~\ref{sec:disk_constancy} we show that the broadband \textit{AstroSat} data place a much weaker constraint on disk-temperature variation and favor a disk that is constant in temperature.}

Our phase-resolved spectral analysis shows a cyclic evolution of the inner accretion disk in GRS~1915+105 during $\rho$-class variability (Figures~\ref{fig:phase_classification} and \ref{fig:hid_xrt}; Table~\ref{tab:xrt_params}). The systematic anti-correlation between inner disk temperature ($T_{\rm in}$) and apparent radius ($R_{\rm in}$) throughout the five phases is consistent with a radiation-pressure-driven instability cycle, as originally proposed by \citet{Neilsen_2011} and predicted theoretically by \citet{LightmanEardley1974} and \citet{ShakuraSunyaev1976}.

During Phases~1--2, the inner disk is moderately extended ($R_{\rm in} \sim 22$--28\,km) and cooling ($T_{\rm in}$ decreasing from $\sim$1.69 to $\sim$1.59\,keV) as radiation pressure inflates the disk outward. The progressive expansion reduces the viscous dissipation rate at larger radii, causing the temperature to drop while the luminosity remains near-Eddington \citep{shakura1973black, LightmanEardley1974, abramowicz1988slim}. By Phase~3, the disk reaches its maximum extent ($R_{\rm in} \sim 37.6$\,km) at the lowest temperature ($T_{\rm in} \sim 1.53$\,keV), marking the turning point where radiation-pressure support can no longer sustain the inflated configuration. The 1--10\,keV flux at this phase ($3.5 \times 10^{-8}$\,erg\,cm$^{-2}$\,s$^{-1}$) is comparable to the burst peak ($3.9 \times 10^{-8}$\,erg\,cm$^{-2}$\,s$^{-1}$), because the larger $R_{\rm in}^2$ compensates for the lower $T_{\rm in}^4$, consistent with a near-constant accretion rate driving the cycle \citep{frank2002accretion}. In Phase~4, the disk collapses inward \citep{abramowicz1988slim, Neilsen_2011}, producing the burst peak at the highest temperature ($T_{\rm in} \sim 1.91$\,keV) and a compact radius ($R_{\rm in} \sim 23.4$\,km). In Phase~5, the post-burst disk contracts further ($R_{\rm in} \sim 18.1$\,km, $T_{\rm in} \sim 1.89$\,keV) as radiation pressure subsides and gravity draws the inner edge inward, resetting conditions for the next expansion cycle \citep{abramowicz1988slim}.

This cyclic behavior is broadly consistent with the predictions of radiation-pressure instability models \citep{abramowicz1988slim, Watarai2001} and aligns with previous observations by \citet{Neilsen_2011} and \citet{Zoghbi2016}. The sequence of expansion, collapse, burst, and contraction follows the canonical heartbeat mechanism established in the literature.

To place these results in the context of the Eddington limit, we estimate the X-ray luminosity from our spectral fits. The observed 1--10\,keV flux ranges from $\sim$$1.8 \times 10^{-8}$ to $\sim$$3.9 \times 10^{-8}$\,erg\,cm$^{-2}$\,s$^{-1}$ across the five phases (Table~\ref{tab:xrt_params}). Adopting a distance of 8.6\,kpc \citep{reid2014grs}, the corresponding 1--10\,keV luminosity is $L_{\rm X} \approx (1.6$--$3.4) \times 10^{38}$\,erg\,s$^{-1}$. For a 12.4\,$M_\odot$ black hole, $L_{\rm Edd} \approx 1.6 \times 10^{39}$\,erg\,s$^{-1}$, so the observed 1--10\,keV luminosity alone corresponds to $\sim$0.1--0.2\,$L_{\rm Edd}$. This 1--10\,keV measurement represents a lower bound on the bolometric luminosity, as our broadband \textit{AstroSat} fits reveal extended Comptonized emission across the full 0.8--30\,keV band. Independent studies have firmly established that GRS~1915+105 persistently accretes at near-Eddington rates ($L \gtrsim 0.3\,L_{\rm Edd}$; \citealt{done2007xrb, remillard2006xrb, McClintock2006}), placing the source well within the regime where radiation-pressure instabilities are expected \citep{abramowicz1988slim}.

\subsection{Disk--corona coupling and the coronal temperature evolution}
\label{sec:diskcorona}

The broadband \textit{AstroSat} analysis reveals a systematic evolution of the coronal electron temperature $kT_{\rm e}$ across the $\rho$ cycle. During Phases~1--4, as $R_{\rm in}$ increases the corona heats progressively: $kT_{\rm e}$ rises from $\sim$6\,keV in Phase~1 through $\sim$10\,keV in Phases~2 and~3 to a maximum of $\sim$14.5\,keV at the burst peak (Phase~4) (Table~\ref{tab:astrosat_params}). This trend reflects the seed-photon starvation mechanism \citep{HaardtMaraschi1991, done2007xrb}: as the inner disk recedes, it subtends a smaller solid angle as seen from the corona, reducing the supply of soft seed photons available for Compton cooling, so the coronal plasma heats to a higher equilibrium temperature.

After the burst peak the corona partially cools as the disk contracts back inward. Between Phase~4 and Phase~5, $R_{\rm in}$ decreases from $\sim$22.9 to $\sim$17.4\,km and $kT_{\rm e}$ drops from its maximum of $\sim$14.5\,keV to $\sim$11.5\,keV, consistent with the partial recovery of seed-photon flux as the disk re-approaches the corona. The cycle then resets to the cool-corona Phase~1 conditions ($kT_{\rm e} \sim 6$\,keV).

The Comptonization photon index $\Gamma$ tracks this evolution \citep{SunyaevTitarchuk1980, Zdziarski1996}: $\Gamma$ steepens from $\sim$1.77 in Phase~1, where the cool, optically thicker corona produces a harder Comptonized spectrum, to $\sim$2.01 at the burst peak (Phase~4), where the hotter and more optically thin corona yields a steeper continuum, and remains close to $\sim$2.00 in Phase~5 before the cycle resets.

In the soft X-ray bands (\textit{Swift} XRT and \textit{AstroSat} SXT), Phase~3 shows the softest spectrum, consistent with a substantially expanded disk. The \textit{AstroSat} LAXPC data show, however, that Phase~3 is not the hardest above $\sim$10\,keV; the hardest hard-band spectrum is that of the low-intensity Phase~1, consistent with its flattest photon index ($\Gamma \sim 1.8$; Table~\ref{tab:astrosat_params}). This energy dependence is consistent with the relative contributions of thermal disk emission and Comptonized continuum \citep{done2007xrb}: at soft X-ray energies the thermal disk dominates, while at hard X-ray energies the Comptonized continuum carries the bulk of the flux.

\subsection{\new{A constant inner disk temperature and the role of the corona}}
\label{sec:disk_constancy}

{
The joint analysis presented in Section~\ref{sec:tin_constancy} shows that the broadband \textit{AstroSat} data are statistically consistent with a single global inner disk temperature, $T_{\rm in} = 1.275 \pm 0.020$\,keV, throughout the $\rho$ cycle, while a fully constant disk (both $T_{\rm in}$ and the apparent emitting area tied) is rejected at high significance. The dominant cycle-to-cycle variability is therefore carried by the Comptonizing corona: in the joint $T_{\rm in}$-tied fit (Table~\ref{tab:joint_constancy}), $kT_{\rm e}$ rises from $\sim$6.5\,keV in Phase~1 to $\sim$13.8\,keV in Phase~5 and the photon index $\Gamma$ steepens monotonically from $\sim$1.82 to $\sim$2.06. The inner disk emitting area also varies, but at a more modest level of $\sim$20\% in apparent $R_{\rm in}$.

This picture has a direct precedent in the literature on GRS~1915+105. Using simultaneous radio and X-ray observations of the $\beta$-class soft X-ray dips, \citet{Vadawale2001} showed that the Comptonized component is the variable spectral element: the low-frequency quasi-periodic oscillation present outside the dip disappears coincidentally with it, while the underlying disk continuum remains comparatively stable, leading the authors to argue that the corona is partially ejected during the dip and provides the radio outflow (see also \citealt{Nandi2001}). The mechanism proposed for the $\rho$-class is different in detail: the heartbeat is conventionally interpreted as a radiation-pressure-driven thermal limit cycle of the inner disk \citep{LightmanEardley1974, ShakuraSunyaev1976, Neilsen_2011, abramowicz1988slim} and not a discrete coronal-ejection event. Despite the difference in driving mechanism, the higher-level conclusion that emerges from our broadband fits, namely that the corona is the dynamic spectral component while the disk temperature stays approximately constant, is the same as in the framework articulated by \citet{Vadawale2001} for $\beta$-class behavior. The thermal-cycle and ejection scenarios are not mutually exclusive; both can be accommodated within a picture in which the corona carries the bulk of the observable spectral variability across multiple variability classes.

The residual $\sim$20\% variation in $R_{\rm in}$ that survives in our $T_{\rm in}$-tied joint fit most likely arises from a combination of effects with no single clean geometric interpretation. A constant-$T_{\rm in}$ disk with a modest, phase-dependent change in the spectral hardening factor $f_c$ produces apparent $R_{\rm in}$ excursions of order $\sim$20\% if $f_c$ varies by $\sim$10\%, which is well within the range expected at near-Eddington luminosities \citep{Davis2019, ShimuraTakahara1995}; the same effect is seen directly in GRS~1915+105 by \citet{Zoghbi2016}, who found the reflection-inferred inner radius nearly constant (varying by $\lesssim$5\%) while the \texttt{diskbb} radius varied by $\sim$30\% across the cycle. Partial coronal covering of the inner disk could further contribute to the apparent normalization swing. Our broadband fits do not resolve which of these mechanisms dominates; what the data do require is that the inner edge does not physically move by $\sim$5\,km across the cycle.

The constant-$T_{\rm in}$ interpretation does not eliminate the role of the radiation-pressure instability framework for the $\rho$-cycle. The cycle still requires a physical driver to modulate the coronal seed-photon supply, and a near-Eddington thermal limit cycle that produces small ($\sim$20\%) excursions in disk emitting area while keeping the temperature pinned by the radiation-pressure equilibrium remains a plausible candidate \citep{abramowicz1988slim, Watarai2001}. In the broadband picture the Comptonizing corona is the principal dynamical component, and the apparent disk-temperature swings inferred from narrow-band fits overstate the underlying disk evolution.
}

\subsection{Cross-instrument differences and the apparent inner radius}
\label{sec:rin_discrepancy}

\new{The narrow-band \textit{Swift} XRT fits return substantially larger apparent disk variations than the broadband \textit{AstroSat} fits, and we attribute this to the limited energy coverage of XRT. The XRT model (\texttt{diskbb + bremss + powerlaw}) does not include an explicit Comptonization component, because the high-energy rollover of any underlying Comptonized continuum, set by $kT_{\rm e}$ which spans $\sim$6--14\,keV across the cycle, lies at or beyond the upper edge of the 1--10\,keV band. Coronal Comptonization is therefore poorly constrained in XRT-only fits; the bremsstrahlung term acts as a low-temperature stand-in for the soft tail of the Comptonized continuum (discussed further below), and the disk normalization and temperature shift to absorb residual coronal variability. The result is apparent $T_{\rm in}$ and $R_{\rm in}$ swings in the XRT block of Table~\ref{tab:xrt_params} that are larger than the underlying physical changes. The broadband \textit{AstroSat} fits constrain the corona directly through LAXPC out to 30\,keV, allowing the disk parameters to settle onto a near-constant temperature.}

A similar model-dependent offset is seen in the best-fit $N_{\rm H}$ values: \textit{Swift} XRT returns $7.2$--$8.6 \times 10^{22}$\,cm$^{-2}$ while \textit{AstroSat} returns $3.4$--$3.9 \times 10^{22}$\,cm$^{-2}$. The SXT extends to lower energies (0.8\,keV) than XRT and constrains the soft X-ray absorption more directly, while the narrower XRT band makes $N_{\rm H}$ degenerate with the disk normalization and the bremsstrahlung component. Previous studies found $N_{\rm H} \sim 5 \times 10^{22}$\,cm$^{-2}$ \citep{Neilsen_2011} and $\sim$4.9--6.1$\times 10^{22}$\,cm$^{-2}$ \citep{Zoghbi2016, Miller2013} with broader-band instruments, intermediate between our two values and more consistent with the \textit{AstroSat} measurement. The phase-to-phase variation within each instrument most likely reflects parameter degeneracies rather than genuine column density changes on the timescales of the $\rho$ cycle.

\label{sec:bremss}
The use of different spectral models for \textit{Swift} XRT (\texttt{diskbb + bremss + powerlaw}) and \textit{AstroSat} (\texttt{diskbb + nthcomp}) also deserves comment. A soft excess is present in XRT-only and SXT-only fits that requires a bremsstrahlung-like component to achieve acceptable $\chi^2$ values. However, when SXT is combined with LAXPC to provide broadband coverage, the \texttt{nthcomp} Comptonization model alone accounts for both the soft excess and the hard X-ray emission. This indicates that the soft excess is the low-energy tail of the Comptonized continuum, not a physically distinct emission component \citep{done2007xrb, Zdziarski1996}, which only becomes apparent as a separate feature when the spectral coverage is insufficient to constrain the full Comptonization model. The convergence onto a single \texttt{nthcomp} component in the broadband fits reinforces the value of broadband coverage in characterizing the spectral components of $\rho$-class sources.

\label{sec:rin_physical}
The apparent inner radii of $R_{\rm in} \approx 17$--38\,km obtained from \texttt{diskbb} fits are consistent with accretion extending to or near the ISCO of GRS~1915+105 once standard corrections are applied. The \texttt{diskbb} normalization yields an apparent radius that must be corrected for spectral hardening and the zero-torque inner boundary condition following \citet{Kubota1998}: the combined correction factor is $\kappa = \xi f_c^2$, where $\xi = 0.412$ accounts for the boundary condition offset and $f_c$ is the spectral hardening (color correction) factor. The canonical value $f_c \approx 1.7$ \citep{ShimuraTakahara1995} gives $\kappa \approx 1.19$; however, at the near-Eddington luminosities characteristic of GRS~1915+105, detailed atmosphere modeling indicates that $f_c$ can approach $\sim$2.0 \citep{Davis2019}, which raises the correction factor to $\kappa \approx 1.65$ and brings our apparent radii into the range $\sim$28--63\,km ($\sim$1.5--3.4\,$R_g$ for $M = 12.4\,M_\odot$; \citealt{reid2014grs}). These corrected values are broadly consistent with the ISCO of GRS~1915+105 ($R_{\rm ISCO} \approx 1.24$--$1.6\,R_g$, or $\sim$23--30\,km, depending on the precise spin within the near-extreme range $a_* > 0.98$ measured by \citealt{McClintock2006}). At near-Eddington accretion rates the disk also enters the slim-disk regime \citep{abramowicz1988slim, Watarai2000}, in which the radial temperature profile flattens from $T \propto r^{-3/4}$ toward $T \propto r^{-1/2}$; \texttt{diskbb} assumes the steeper thin-disk profile and therefore systematically underestimates the emitting area, which can place the uncorrected $R_{\rm in}$ values below the true ISCO even when the disk physically extends to it.

\subsection{Comparison with previous studies}
\label{sec:comparison}

Our results are broadly consistent with the seminal work of \citet{Neilsen_2011}, who first performed detailed phase-resolved spectral analysis of the $\rho$ variability using simultaneous \textit{Chandra} HETGS and \textit{RXTE} data. \citet{Neilsen_2011} found inner disk radii of $R_{\rm in} \sim 70$--110\,km using the \texttt{ezdiskbb} model with a color correction factor $f = 1.9$, compared to our \textit{Swift} XRT values of $R_{\rm in} \sim 18$--38\,km obtained with \texttt{diskbb}. This systematic offset arises primarily from the different disk models employed: \texttt{ezdiskbb} incorporates a zero-torque inner boundary condition that tends to yield larger apparent radii than \texttt{diskbb} for comparable data \citep{Kubota1998, Zimmerman2005}. We extend the work of \citet{Neilsen_2011} in two ways. We utilize 24 \textit{Swift} XRT observations spanning 2014--2016, a substantially larger statistical sample than the single-epoch \textit{Chandra}/\textit{RXTE} study, and the consistency of the spectral evolution across multiple $\rho$-class episodes demonstrates the robustness and repeatability of the phase pattern over years. The simultaneous \textit{AstroSat} SXT+LAXPC coverage (0.8--30\,keV) also provides independent constraints on both the thermal disk and Comptonized continuum, complementing the earlier \textit{Chandra}/\textit{RXTE} analysis of \citet{Neilsen_2011} and the \textit{NuSTAR}/\textit{Chandra} work of \citet{Zoghbi2016}, with the broad LAXPC bandpass in particular enabling the joint $T_{\rm in}$-tied test described in Section~\ref{sec:tin_constancy}. A similar phase-resolved and timing approach has recently been applied to another black hole binary by \citet{Patel2025}, who tracked a continuous transition from type-C quasi-periodic oscillations to a heartbeat state in 4U~1630--47, establishing it as a second system, alongside GRS~1915+105, in which this transition is seen.

\section{Conclusions}
\label{sec:conclusions}

We have presented a phase-resolved spectral and timing analysis of the $\rho$-class (``heartbeat'') variability in the Galactic microquasar GRS~1915+105 using 24 \textit{Swift} XRT observations spanning 2014--2016 and two broadband \textit{AstroSat} LAXPC/SXT observations. Our main conclusions are:

\begin{itemize}
\item Phase-resolved \textit{Swift} XRT spectroscopy shows an apparent $T_{\rm in}$--$R_{\rm in}$ anti-correlation across the $\rho$ cycle (the inner disk appears to expand from $\sim$22 to $\sim$38\,km while cooling from $\sim$1.7 to $\sim$1.5\,keV in Phases~1--3, then contract to $\sim$18\,km post-burst), repeatable across all 24 observations spanning 2014--2016. \new{The broadband \textit{AstroSat} data, however, are statistically consistent with a single constant inner disk temperature, $T_{\rm in} = 1.275 \pm 0.020$\,keV (joint $T_{\rm in}$-tied fit at $\chi^2_\nu = 1.003$ for 1881 dof; $\Delta\chi^2 = +4.3$ for 4 added constraints). A fully constant disk is rejected ($\Delta\chi^2 = +175.9$ when both $T_{\rm in}$ and the disk normalization are tied), with apparent $R_{\rm in}$ varying by $\sim$20\% between $18.1 \pm 0.7$\,km (Phase~5) and $21.9 \pm 0.7$\,km (Phase~4). The disk is thus constant in temperature but modestly variable in apparent emitting area through the cycle.}

\item Broadband \textit{AstroSat} SXT+LAXPC data show a systematic correlation between disk extent and coronal electron temperature ($kT_{\rm e}$ rising from $\sim$6 to $\sim$14.5\,keV during disk expansion through Phases~1--4, partially recovering to $\sim$11.5\,keV post-burst in Phase~5), driven by geometric modulation of the seed-photon supply via the Haardt--Maraschi mechanism \citep{HaardtMaraschi1991}. Energy-dependent spectral hysteresis in the HID and CCD confirms this disk-corona feedback, with Phase~3 appearing softest in XRT/SXT while the hardness ordering changes in LAXPC, where the low-intensity Phase~1 is the hardest above $\sim$10\,keV. \new{The spectral variability through the cycle is therefore driven mainly by the corona, consistent with the framework established for GRS~1915+105 by \citet{Vadawale2001}; the larger apparent $T_{\rm in}$ swings inferred from \textit{Swift} XRT alone are a consequence of its limited 1--10\,keV bandpass, and broadband coverage to $\sim$30\,keV is needed for an unbiased disk--corona decomposition.}

\item These results provide new observational constraints on near-Eddington accretion physics relevant to ULXs, tidal disruption events, and AGN variability \citep{Kaaret2017, Komossa2015, Czerny2009}. The broad spectral range of our \textit{AstroSat} SXT+LAXPC analysis (0.8--30\,keV) enables us to simultaneously probe both the thermal disk and the Comptonized corona, directly revealing the disk-corona coupling and its evolution across the $\rho$ cycle, a result inaccessible to narrow-band observations. The disk-corona feedback mechanism demonstrated here provides an observational constraint for theoretical models of near-Eddington systems, and our phase-resolved continuum analysis provides complementary constraints on the disk structure that drives the wind variability reported by \citet{Zoghbi2016}.
\end{itemize}

In the future, the combination of high-resolution grating spectroscopy and broadband timing available with \textit{XRISM} \citep{Tashiro2018} will enable simultaneous constraints on the disk wind, reflection spectrum, and continuum across the $\rho$ cycle, directly testing whether the apparent $R_{\rm in}$ variations reflect true disk motion or color-correction effects \citep{Zoghbi2016, ShimuraTakahara1995}. The next generation of X-ray observatories, including \textit{Athena} \citep{Nandra2013}, will extend this to fainter systems and higher redshifts, placing GRS~1915+105 in the broader context of accretion variability across the mass scale.

An important caveat on the physical interpretation of the apparent $R_{\rm in}$ evolution comes from \citet{Zoghbi2016}, who used simultaneous \textit{NuSTAR} and \textit{Chandra} data to compare the inner radius inferred from the disk blackbody ($R_{\rm bb}$) with that from the relativistic reflection spectrum ($R_{\rm ref}$). They found the disk-blackbody radius $R_{\rm bb}$ to be far more variable across the $\rho$ cycle than the reflection-inferred radius $R_{\rm ref}$, and concluded that the apparent radius changes in blackbody-based models may partly reflect variations in the disk color correction factor $f_c$ \citep{ShimuraTakahara1995} driven by changes in disk composition and opacity, and not true physical motion of the disk inner edge. Our \textit{Swift} XRT and \textit{AstroSat} analyses are based solely on the disk continuum and cannot directly test this. The qualitative anti-correlation we observe is robust, but the absolute amplitude of the $R_{\rm in}$ variations should be interpreted with this caveat in mind. Future observations capable of simultaneously resolving both the disk continuum and the relativistic reflection spectrum (e.g., \textit{XRISM}, \citealt{Tashiro2018}; or \textit{Athena}, \citealt{Nandra2013}) will be essential to disentangle these effects.

\section*{Acknowledgments}

We thank the \textit{Swift} and \textit{AstroSat} teams for scheduling and executing these observations. We acknowledge the use of public data from the \textit{AstroSat} archive at the Indian Space Science Data Center (ISSDC). We thank the SXT and LAXPC Payload Operation Centers at TIFR, Mumbai, for verifying and releasing the data. We also thank the respective instrument teams for their assistance with the data reduction and analysis. This research has made use of data obtained from the High Energy Astrophysics Science Archive Research Center (HEASARC), provided by NASA's Goddard Space Flight Center.

\bibliographystyle{plainnat}
\bibliography{ref}

\end{document}